\documentclass[iop]{emulateapj}

\usepackage{paralist}
\usepackage{graphics,graphicx}
\usepackage{apjfonts}

\begin{document}

\shortauthors{Rappaport et al.}
\shorttitle{Triple-Star Candidates Among the {\em Kepler} Binaries}
\title{Triple-Star Candidates Among the {\em Kepler} Binaries}

\author{S. Rappaport\altaffilmark{1}, K. Deck\altaffilmark{1},  A. Levine\altaffilmark{2}, T. Borkovits\altaffilmark{3}, J. Carter\altaffilmark{4,5}, I. El Mellah\altaffilmark{6}, R. Sanchis-Ojeda\altaffilmark{1}, B. Kalomeni\altaffilmark{7} }  

\altaffiltext{1}{M.I.T. Department of Physics and Kavli
Institute for Astrophysics and Space Research, 70 Vassar St.,
Cambridge, MA, 02139; sar@mit.edu, kdeck@mit.edu, rsanchis@mit.edu} 
\altaffiltext{2}{37-575 M.I.T. Kavli
Institute for Astrophysics and Space Research, 70 Vassar St.,
Cambridge, MA, 02139; aml@space.mit.edu} 
\altaffiltext{3}{Baja Astronomical Observatory, H-6500 Baja, Szegedi \'ut, Kt. 766, Hungary; Konkoly Observatory, MTA CSFK, H-1121 Budapest, Konkoly Thege M. \'ut 15-17, Hungary; ELTE Gothard-Lend\"ulet Research Group, H-9700 Szombathely, Szent Imre herceg \'ut 112, Hungary; borko@electra.bajaobs.hu} 
\altaffiltext{4}{Harvard-Smithsonian Center for Astrophysics, 60 Garden Street Cambridge, MA 02138 USA, e-mail: jacarter@cfa.harvard.edu} 
\altaffiltext{5}{Hubble Fellow}
\altaffiltext{6}{ENS Cachan, 61 avenue du Pr\'esident Wilson, 94235 Cachan, France; ielmelah@ens-cachan.fr}
\altaffiltext{7}{Department of Astronomy and Space Sciences, University of Ege, 35100 Bornova-Izmir, Turkey; Department of Physics, Izmir Institute of Technology, Gulbahce, Urla 35430 Izmir, Turkey}

\begin{abstract}
We present the results of a search through the photometric database of {\em Kepler} eclipsing binaries (Pr\v{s}a et al.~2011; Slawson et al.~2011) looking for evidence of hierarchical triple star systems. The presence of a third star orbiting the binary can be inferred from eclipse timing variations. We apply a simple algorithm in an automated determination of the eclipse times for all 2157 binaries. The ``calculated'' eclipse times, based on a constant period model, are subtracted from those observed. The resulting $O-C$ (observed minus calculated times) curves are then visually inspected for periodicities in order to find triple star candidates.  After eliminating false positives due to the beat frequency between the $\sim$1/2-hour {\em Kepler} cadence and the binary period, 39 candidate triple systems were identified.  The periodic $O-C$ curves for these candidates were then fit for contributions from both the classical Roemer delay and so-called ``physical'' delay, in an attempt to extract a number of the system parameters of the triple. We discuss the limitations of the information that can be inferred from these $O-C$ curves without further supplemental input, e.g., ground-based spectroscopy. Based on the limited range of orbital periods for the triple star systems to which this search is sensitive, we can extrapolate to estimate that at least 20\% of all close binaries have tertiary companions.

\end{abstract}


\section{Introduction} 
\label{sec:intro}

Triple star systems are appealing objects for study for a number of reasons. The orbital architecture and masses of the constituent stars can inform us about the not-so-well understood process of the formation of systems of multiple stars (see, e.g., Boss 1991; 1995; Bodenheimer et al.~2000; Sterzik, Tokovinin, \& Shatsky 2003; Bate 2009; Reipurth \& Mikkola 2012). As one example, it is known that close binary systems cannot have formed in their current configurations; during their protostellar phase the stellar radii would have been much too large to fit inside their current orbits. The presence of an orbiting third star in the system could provide a natural mechanism, through Kozai cycles (Kozai 1962) with tidal friction, for the initially wide binary to lose angular momentum and become close (Kiseleva, Eggleton, \& Mikkola 1998; Eggleton \& Kiseleva-Eggleton 2001; Fabrycky \& Tremaine 2007). This mechanism has also been proposed as a way to explain the blue-straggler stars found predominantly in globular clusters (Perets \& Fabrycky 2009). The orbital architecture of a triple star system can also in principle inform us about the final contraction of the interstellar cloud that formed the system, provided the dynamical evolution of the system has left the initial configuration relatively unaltered (see, e.g., Boss 1991; Bate 2009; Reipurth \& Mikkola 2012).

Moreover, understanding the relative frequency of binaries vs.~triples and quadruples (see, e.g., Tokovinin et al.~2006; Pribulla \& Rucinski 2006; Raghavan et al.~2010) is important in anticipating what other unseen stars in any particular system may be present. The hypothetical presence of such bodies may be important in explaining various effects that are observed in these binaries, but not otherwise explained (see, e.g., Eggleton \& Kiseleva-Eggleton 2001, and references therein).  Finally, while studies of binary star evolution, and especially the phases involving mass transfer, have dramatically transformed our overall understanding of stellar evolution and the exotic remnants, such as binary neutron stars, that are left in the late phases, studies of the little-explored triple star evolution promise to involve yet several more layers of complexity.

There are at least five ways of finding triple star systems.  These include (i) visually resolving bound star systems, including with adaptive optics and optical/IR interferometry (see, e.g., Tokovinin et al.~2006; Rucinski, Pribulla, \& van Kerkwijk 2007; Raghavan et al.~2010).  (ii)  Observing the presence of three different stellar spectra in an apparently single object provides an excellent starting point for the discovery of triples (see, e.g., Zucker, Torres, \& Mazeh 1995; D'Angelo, van Kerkwijk, \& Rucinski 2006).  (iii) Doppler spectroscopy (i.e., measurements of radial velocity) carried out over intervals at least as long as the binary period in the system, and a substantial portion of the period of the triple, is the most informative (see, e.g., Carter et al.~2011).  (iv) Direct observations of eclipses by all three bodies is also exceptionally interesting, but such systems are relatively rare (see, e.g., Carter et al.~2011; Derekas et al.~2011; Carter et al.~2013).  Finally, as has been done for more than a century (v) long-term timing of binary eclipses can reveal periodic perturbations to the otherwise linear progression of eclipse times with cycle number (see, e.g., Irwin 1952; Fabrycky 2010; Steffen et al.~2011; Gies et al.~2012; Borkovits et al.~2013).  It is the latter approach which is the subject of this paper.  We also note that this method of timing variations has been used to great success in measuring orbits and masses of multi-planet systems (see, e.g., Holman et al.~2010; Lissauer et al.~2011; Carter et al.~2012), though the mass and period ratios of the perturbers are different in planetary systems vs.~triple star systems.

However, each of these methods suffers from some limitations, and each probes different regimes in the ratio of the binary period to that of the triple systems.   In the case of timing binary eclipses, this can be done quite accurately from ground-based measurements, at least on bright objects, and such studies have provided substantial hints of the presence of third bodies (see, e.g., Pribulla \& Rucinski 2006).  The difficulty here has been that ground-based eclipse timing studies are subject to frequent interruptions due to the diurnal, lunar, and seasonal cycles, not to mention the weather.  In this work we make use of three years of nearly continuous observations by {\em Kepler} of some 2000 eclipsing binaries to identify candidates for triple star systems.   

The {\em Kepler} mission (Borucki et al.~2010; Koch et al.~2010; Caldwell et al.~2010) has been observing some 157,000 stars, including $\sim$2000 eclipsing binaries, for the past three years. The continuous monitoring of these eclipsing systems, in combination with the exquisite high photometric precision of the {\em Kepler} mission (Jenkins et al.~2010a; 2010b), is unprecedented in the history of observational astronomy.  As a result, this photometric data set of eclipsing binaries is able to make a serious contribution to the endeavor of identifying promising triple star candidates for followup studies of radial velocity via Doppler spectroscopy.  Already, the {\em Kepler} observations have yielded some five triple star systems identified directly by third-body eclipses of the binary (Carter et al.~2011; Derekas et al.~2011; Slawson et al.~2011) while a number of others have been inferred to be triples by evidence for systematic eclipse timing variations (``ETVs'') of binaries (Fabrycky 2010; Slawson et al.~2011; Steffen et al.~2011; Carter et al.~2013). The Slawson et al.~(2011) catalog of binaries, in which ten of these triples are briefly mentioned, was based on only 120 days of {\em Kepler} data, whereas approximately an order of magnitude more data now exist.  

In this study we present the results of a comprehensive search of the {\em Kepler} data base of binary systems for evidence of the presence of a third star. This was done by searching for periodic features in so-called $O-C$ curves (observed minus calculated eclipse times) of some 2000 eclipsing binaries. We find 39 good candidates for triple stars.   In addition to exhibiting the periodic variations in the $O-C$ curves indicative of a triple system, several of our candidates feature additional evidence for being triple. For example, two of the systems have third-body eclipses, while seven of them exhibit secular variations in the depths of the binary eclipses indicative of precession of the orbital plane of the binary. As we show, 19 of the systems exhibit dominant classical Roemer delays, while another 11 have dominant physical delays (due to perturbations to the binary ``clock'', i.e., its orbital eclipse period). The especially interesting feature of these candidates is that we can directly follow perturbations to the binary orbit and/or the classical Roemer delay {\em continuously} over several cycles of the triple.

The processing of the {\em Kepler} data for the 2157 eclipsing binaries is described in \S \ref{sec:prep}. Production of an $O-C$ curve for each system is discussed in \S \ref{sec:ecltime}, while an overview of our triple star candidates is presented in \S \ref{sec:triples}.  Expressions for the various effects that appear in the $O-C$ curves are given quantitatively in \S \ref{sec:OmCeffects}.  Our approach to the analysis of the $O-C$ curves, in order to extract as much information about the physical system parameters as possible, is described in \S \ref{sec:analysis}.  Our results for the 39 triples found in the search are presented in \S \ref{sec:results}.  We discuss the limitations on the determination of system parameters using only the {\em Kepler} eclipse timing data, without supplemental information that could be provided by ground-based spectral observations (and in some cases by the {\em Kepler} data themselves). All of these systems will require such follow-up observations in order to definitively determine the masses of the three stars and the orbital elements. In \S \ref{sec:discuss} we discuss our results, with emphasis on what can be learned from only the $O-C$ curves.  Finally, we attempt to estimate the fraction of close binaries with tertiary stars of orbital periods $\lesssim$ few years.

\section{Data preparation}
\label{sec:prep}

\subsection{{\em Kepler} binary data set}

The data we use for this study are long-cadence (LC) lightcurves for all
binaries published in the latest {\em Kepler} eclipsing binary catalog (Slawson et al.~2011; see also Pr\v{s}a et al.~2011). 
We used all the files from Quarter 1 through
Quarter 13 which were available for retrieval from the Multimission Archive at
STScI (MAST). 
The data used had all been reprocessed with the PDC-MAP
algorithm (Stumpe et al. 2012; Smith et al. 2012), which removes much
of the instrumental noise from the flux time series while retaining the bulk of the
astrophysical variability in sources. For each quarter, we normalized the
flux series to its median value, and then stitched the quarters together into
a single file for each source. 

\subsection{Filtering the data}

The next step in the data processing was to apply a high-pass filter, based on the known period of the binary system.  We took the stitched 13 quarters of data, described in section 2.1, and filtered out the low frequencies (starspot activity, in particular), in the following way.  First, the data were convolved with a boxcar function of duration equal to the known binary period.  Second, the smoothed data were subtracted from the unsmoothed data.  Frequency components below the frequency of the binary orbit are thereby largely removed, while leaving temporal structures that are shorter than the binary orbital period.  The eclipses themselves are essentially unaffected. 

The reference epoch for all times in this paper is Barycentric Julian Day 2454900.

\section{Eclipse Timing Analysis: $O-C$ Curves}
\label{sec:ecltime}

\subsection{Measuring Eclipse Times}

The baseline algorithm we utilized for determining the eclipse times consists simply of testing each flux point in the {\em Kepler} data set for a local minimum and fitting a parabola to the lowest three points in the local minimum.  Then the fitted parabola is used to interpolate between {\em Kepler} samples to find a more accurate time of eclipse minimum.  As we show, this algorithm is quite good for short orbital period binaries, but begins to lose accuracy for longer-period binaries when the eclipse duration may consist of a substantial number of {\em Kepler} long-cadence samples.  To carry out our initial search for periodic variations in the $O-C$ curves, we used this basic algorithm exclusively.  However, after interesting systems were identified, we recalculated more accurate $O-C$ curves using a better algorithm that involves more of the eclipse profile (T. Borkovits, unpublished) for a handful of the binaries with periods with $P_{\rm bin} \gtrsim 6$ days\footnote{After this work had essentially been completed, we developed a more sophisticated eclipse timing code based on a formal cross-correlation of the epoch-folded binary light curve with the {\em Kepler} data train.  We found all 39 of the triple star candidates with this improved code, including four new candidates that the original search missed.  The quality of the $O-C$ curves was hardly changed for most of the systems with binary period $P_{\rm bin} \lesssim 10$ days, but there were some improvements, i.e., lower scatter, for a few of the longer period systems.  In eight cases, where the $O-C$ curve significantly improved over the simple quadratic fitting algorithm, and where the $O-C$ curve had not already been upgraded using the Borkovits (unpublished) code, we used those $O-C$ results rather than the original.}.

The parabola to be fit is of the form:
\begin{eqnarray}
F_n = \alpha(t_n-\delta t)^2 + F_{\rm min}
\end{eqnarray}
where $n = 1, 2,$ or 3; $t_1 \equiv -1$,  $t_2 \equiv 0$, and $t_3 \equiv +1$; and $\delta t$ is the offset of the time of the minimum with respect to the time of the point with the lowest flux of the three {\em Kepler} samples.  The times are all dimensionless, and are in units of $\Delta t_{\rm LC}$ = 1765.46 sec, the {\em Kepler} long-cadence sampling interval.  We note that the parameter $\alpha$ in this expression  implicitly encompasses information about the relative sizes of the stars, limb darkening, orbital inclination, and so forth.  Presumably for a given binary system this parameter remains a constant, though in practice, effects such as time-varying starspots, can slightly modify $\alpha$.

Since not all binary eclipses are well represented by a simple quadratic function near minimum, we also considered a quartic shape.  This is the next simplest shape for any symmetric eclipse profile.  Because there are four parameters that describe a symmetric quartic, this would require four or more flux points to fit.  Five is the minimum number of points in a symmetric arrangement which can have a lowest flux point with two higher-flux points on either side.  However, we judged this to be too many to use for the shortest period binaries -- in some cases, the eclipse is only a few {\em Kepler} cadence points wide.  Thus, to get a flavor for how a quartic might fit, we utilized a function of the following form: 
\begin{eqnarray}
F_n = \alpha(t_n-\delta t)^2 + \beta \alpha(t_n-\delta t)^4+ F_{\rm min}
\end{eqnarray}
where the parameter $\beta$ was {\em fixed} at a representative value of 0.3.  Thus, there are still only three parameters to fit analytically to three data points.  Again, note that all the times are dimensionless (i.e., in units of $\Delta t_{\rm LC}$).  We also tried other values for $\beta$, but found no improvement (i.e., reduced rms scatter) in the ``quartic'' algorithm.

Once we found a potential eclipse time, and a corresponding value of $F_{\rm min}$ we required that it be less than a certain threshold flux in order to be judged an actual eclipse and not just an uninteresting local minimum in the flux.  Formally, we somewhat arbitrarily required that
\begin{eqnarray}
F_{\rm min} < 0.4 \cdot F_{\rm ecl} + 0.6 
\end{eqnarray}
where $F_{\rm ecl}$ is the flux at the bottom of the primary eclipse in the folded light curve, and recall that the fluxes are all normalized to unity.  In some cases, this allowed the secondary eclipse to also be picked up, but these were distinguished by the $\sim$180$^\circ$ phase shift from the primary eclipse.

In general, the quadratic function produced better results than the quartic, i.e., less scatter in the $O-C$ curves,  but yielded a comparable number of candidate triple stars.  Both functions were equally susceptible to spurious periodicities (see \S \ref{sec:search}).

As a separate piece of the analysis, we also deliberately found the times of the secondary eclipses.  However, in this work we do not directly utilize their $O-C$ curves in the timing analyses.  We do discuss what supplemental information the secondary eclipses can yield in the case of eccentric binaries.  We also tabulate which systems have secondary eclipses whose $O-C$ curves exhibit {\em different} behavior than that of the primary eclipse.  

Finally, we note that even though the nominal separation of the flux points in the long-cadence mode, $\Delta t_{\rm LC}$, is 1765.46 sec, we were able to determine the times of eclipse minima to a typical empirically determined accuracy of $\sim 20 - 100$ sec, or $\lesssim 5$\% of the timing metric.  We list the rms residuals to the model fits for each source among our tabulated results.  

\subsection{Searching for Interesting $O-C$ Curves}
\label{sec:search}

As we search for potential triple-star signatures among the $O-C$ curves, we find many that exhibit spurious periodicities.  These false positives are most often due to a beat between the frequency of the {\em Kepler} cadence and the frequency of the binary orbit.  The two prominent beat frequencies are given by:
\begin{eqnarray}
f_{\rm beat,1}  & = & f_{\rm LC} -  f_{\rm bin} \cdot {\rm int}\left(\frac{f_{\rm LC}}{f_{\rm bin}}\right) \\
f_{\rm beat,2}  & = & f_{\rm bin} \left[{\rm int}\left(\frac{f_{\rm LC}}{f_{\rm bin}}\right) + 1 \right]- f_{\rm LC} 
\label{eqn:beat}
\end{eqnarray}
where ``int'' gives the truncated integer value, and $f_{\rm bin} \equiv 1/P_{\rm bin}$ and $f_{\rm LC} \equiv 1/\Delta t_{\rm LC}$. For each $O-C$ curve that we compute, we display these two prominent expected beat periods.  If there is a match between a predicted beat period and the detected period in the $O-C$ curve, that object is eliminated as a possible triple star candidate.  We note that these beat frequencies change (sometimes fairly obviously) during the course of a year. This is due to the fact that the time of each long-cadence measurement was corrected to the Solar System Barycenter.

As another caveat, we note that many of the contact binaries exhibit a pseudo-random walk in eclipse phase as well as quasi-periodic behavior with typical amplitudes of $\sim$300 sec rms (Tran et al.~2013).  In addition, the $O-C$ curves for the secondary eclipses in these systems are often anti-correlated with the primary $O-C$ curve (Tran et al.~2013).  The characteristic timescales for these cyclic changes in phase can range from weeks to many months.  Therefore, one should be cognizant of the possibility that $O-C$ periods of the order of the 3-year {\em Kepler} data interval might simply be the lowest prominent frequency of a random-walk process -- especially for contact binaries.  In this work we remain mindful of this possibility.  We therefore generally require two full orbital cycles (i.e., with period of the triple system $P_{\rm trip} \lesssim $ 600 days) that are strictly periodic before we are reasonably confident that a binary is also a good triple star candidate.  However, our collection of 39 triple star candidates does contain nine systems with $P_{\rm trip} \gtrsim $ 600 days (six of these have $P_{\rm bin} < 1.1$ day; three are classified as `contact binaries'). The reader can be the judge of the validity of these candidates.  

\begin{figure*}
\begin{center}
\includegraphics[width=0.95\linewidth]{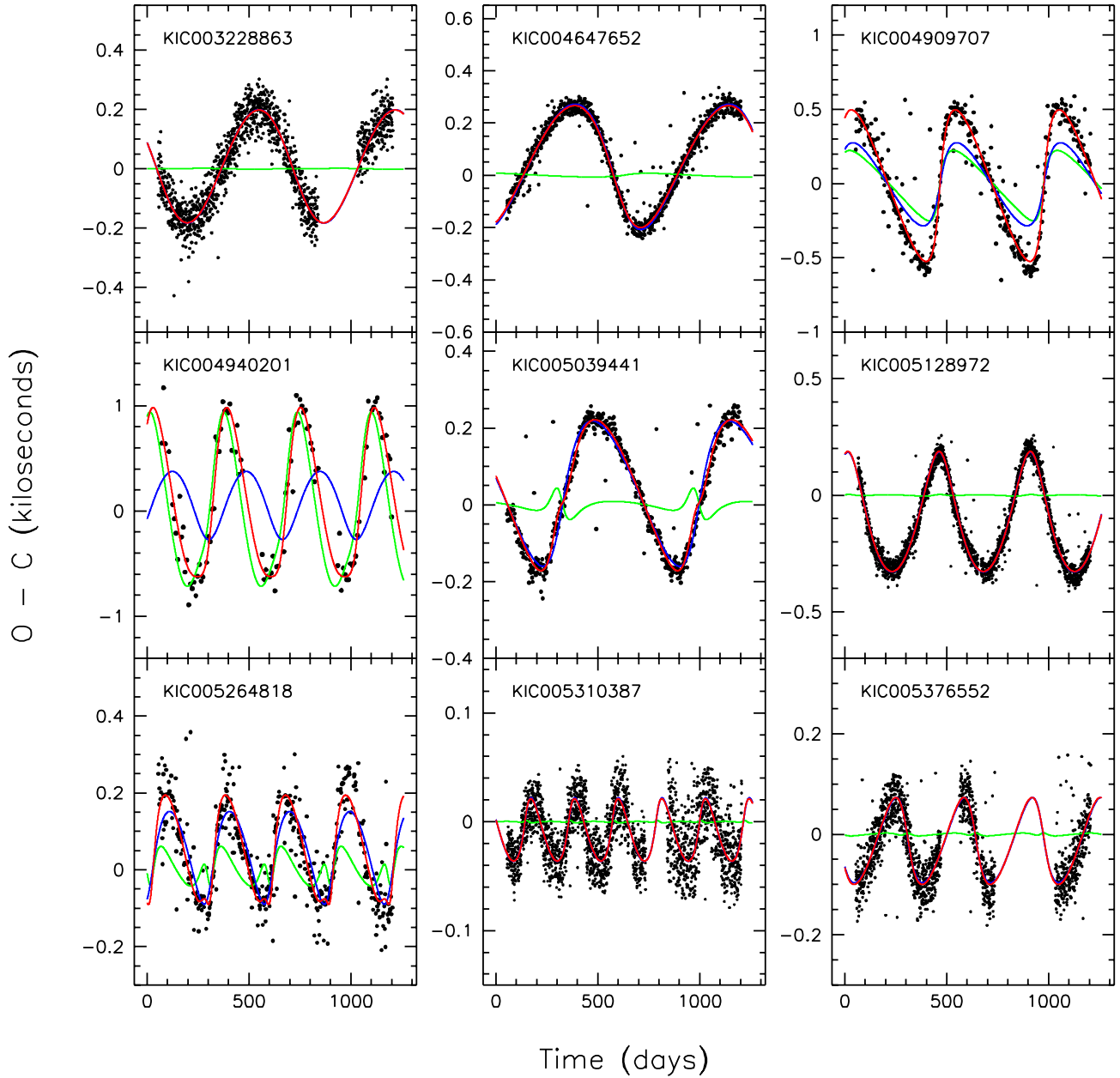}
\caption{$O-C$ data and model fits for 9 systems with KIC numbers between 3228863 and 5376552.  The red curves are the total model $O-C$ values.  Dark blue is the model fit for the Roemer delay (eq.~(\ref{eqn:Roem})).  Light green curves represent the total physical delay (sum of eqs.~(\ref{eqn:phys1}) and (\ref{eqn:phys2})).  Note that the vertical scales are different on all of the plots; the amplitudes of the $O-C$ curves range from a low of 30 sec to a high of 1000 sec.  The linear and quadratic terms in the fit have been subtracted before the plot is made.}
\label{fig:OmC1}
\end{center}
\end{figure*}

\subsection{Candidate Triples}
\label{sec:triples}

After eliminating as many false positives as we were able, we were left with a list of 39 candidate triple star systems with convincing eclipse timing variations (``ETVs'').  The {\em Kepler} Input Catalog (KIC; Batalha et al.~2010) numbers of our 39 candidate triple stars are summarized in Table 1, along with other properties of the targets that are provided in the KIC.  Among other parameters, we list the orbital period of the binary, the {\em Kepler} magnitude ($K_p$) and $T_{\rm eff}$ of the integrated light from the system, the depths of the primary and secondary eclipses, the mass ratio and ``third light'' parameter (as found with the {\em Phoebe} binary light curve emulator; see section \ref{sec:binaryLC}), and an approximate binary orbital eccentricity (taken from the Slawson et al.~2011 catalog).  

\begin{figure*}
\includegraphics[width=0.95\textwidth]{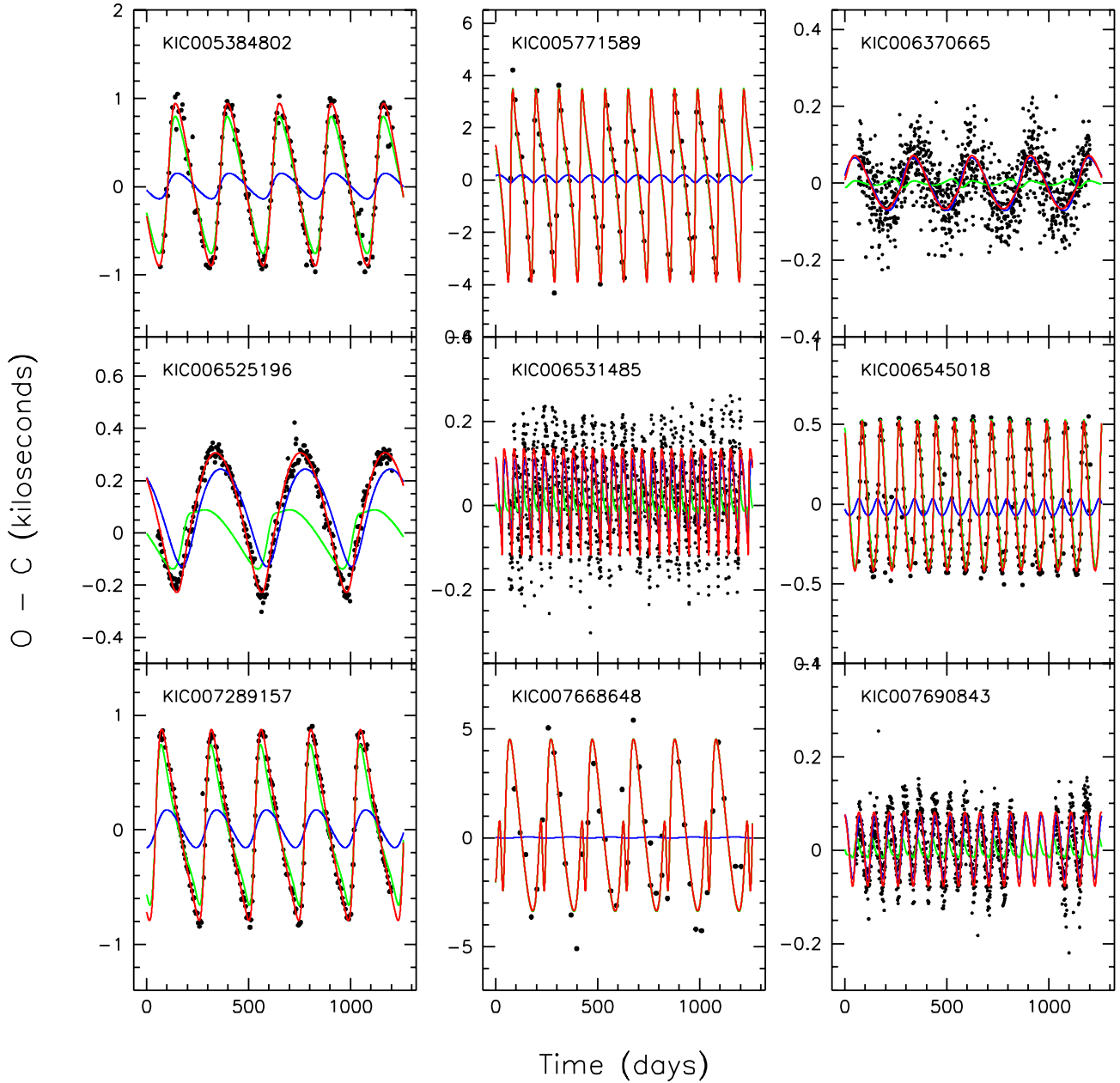}
\caption{$O-C$ data and model fits for 9 systems with KIC numbers between 5384802 and 7690843.  The red curves are the total model $O-C$ values.  Dark blue is the model fit for the Roemer delay (eq.~(\ref{eqn:Roem})).  Light green curves represent the total physical delay (sum of eqs.~(\ref{eqn:phys1}) and (\ref{eqn:phys2})).  Note that the vertical scales are different on all of the plots; the amplitudes of the $O-C$ curves range from a low of 60 sec to a high of 5000 sec.  The linear and quadratic terms in the fit have been subtracted before the plot is made.}
\label{fig:OmC2}
\end{figure*}

\begin{figure*}
\centering
\includegraphics[width=0.95\linewidth]{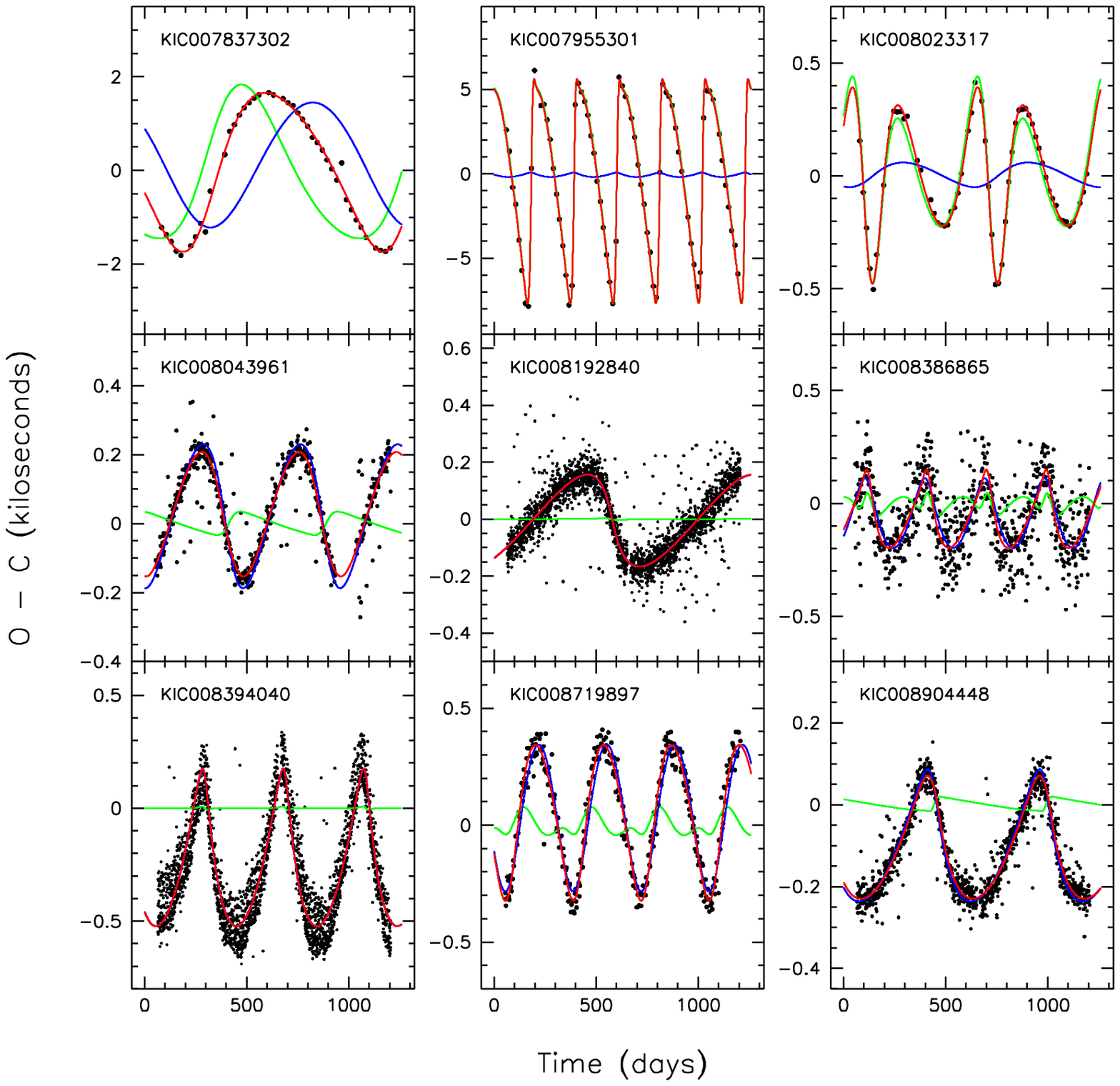}
\caption{$O-C$ data and model fits for 9 systems with KIC numbers between 7837302 and 8904448.  The red curves are the total model $O-C$ values.  Dark blue is the model fit for the Roemer delay (eq.~(\ref{eqn:Roem})).  Light green curves represent the total physical delay (sum of eqs.~(\ref{eqn:phys1}) and (\ref{eqn:phys2})).  Note that the vertical scales are different on all of the plots; the amplitudes of the $O-C$ curves range from a low of $\sim$150 sec to a high of 6000 sec.  The linear and quadratic terms in the fit have been subtracted before the plot is made.}
\label{fig:OmC3}
\end{figure*}

The $O-C$ curves for all 39 of the candidate triple star systems are shown in Figs.~\ref{fig:OmC1}, \ref{fig:OmC2}, \ref{fig:OmC3}, \ref{fig:OmC4}, and \ref{fig:OmC5}.   As the reader will see, there is a great variety of shapes, of amplitudes, and of statistical quality.  These, and formal model fits to them, are discussed in detail in the following sections.  In general, the rms deviations from the best fitting curves are in the range of 20 to 100 sec.  The amplitudes of the $O-C$ curves range from a minimum of 30 sec to a maximum of nearly 6000 sec.  The inferred orbital periods of the triple star systems range from 48 days to 959 days.

\section{Sources of ETV Due to Third Stars} 
\label{sec:OmCeffects}

\subsection{General Expressions}

An eclipsing binary can be thought of as a clock, where the clock ``ticks'' are the binary eclipses. If the binary is circular and isolated in space, then the arrival times of the eclipse events at the solar system barycenter occur at a constant rate -- assuming that the binary orbit is neither decaying nor expanding. When the binary is part of a hierarchical triple system, where both the binary and the third star orbit their common center of mass, the clock ``ticks'' are no longer regular. There are two basic effects that cause these eclipse arrival times to deviate from the pattern of a regular clock, on the timescale of the orbital period of the triple.

In this work we define the ``orbit of the triple system'' (alternatively, ``outer orbit'') as that of an equivalent binary system comprised of the third star and a mass $M_{\rm bin}$ located at the center of mass of the binary system.  Here we have defined $M_{\rm bin}$ as the mass of the inner binary.

\subsubsection{Roemer delay}

The first important effect is the classic Roemer delay (or light travel time delay) that results from the changing projected distance along the line of sight of the center of mass of the binary from the center of mass of the triple star system. 
The expression for the contribution to the $O-C$ curve from the Roemer delay, $\mathcal{R}(t)$, is 
\begin{eqnarray}
\frac{\mathcal{R}(t)}{A_{Roem}} \simeq \left[{(1-e^2)}^{1/2}\sin u \cos \omega + (\cos u-e) \sin \omega \right]
\label{eqn:Roem}
\end{eqnarray}
where $u(t)$ is the eccentric anomaly, $\omega$ the longitude of periastron, and $e$ the eccentricity, all describing the orbit of the triple star system (i.e., the CM of the binary moving about the CM of the triple star system).  The amplitude of the Roemer delay is:
\begin{eqnarray}
A_{\rm Roem} = \frac{G^{1/3}}{c(2\pi)^{2/3}} P_{\rm trip}^{2/3} \left[\frac{M_3 \sin i_{\rm trip}}{M_{\rm trip}^{2/3}}\right]
\label{eqn:Roemamp}
\end{eqnarray}
where $M_3$ is the mass of the third star; $M_{\rm trip}$ is the total mass of the triple star system, i.e., $M_{\rm trip} \equiv M_3+M_{\rm bin}$; $i_{\rm trip}$ is the inclination of the orbital plane of the triple star system with respect to the plane of the sky; and $P_{\rm trip}$ is the orbital period of the triple.

	A diagram showing the triple star system geometry is given in Fig.\,\ref{fig:geometry} (where some of the quantities labeled appear only in the physical delay function -- see below for definitions).

\begin{figure*}
\centering
\includegraphics[width=0.95\linewidth]{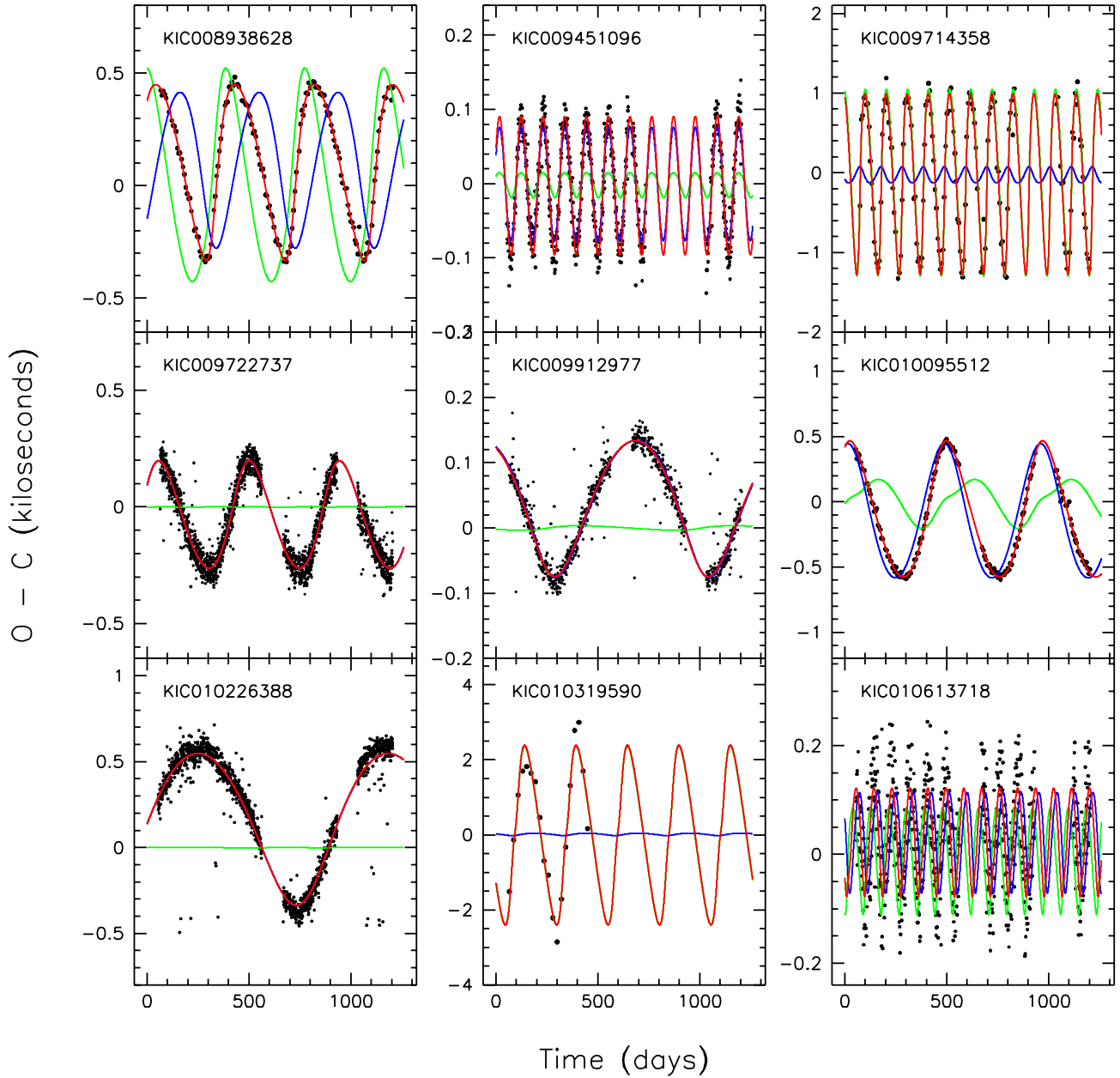}
\caption{$O-C$ data and model fits for 9 systems with KIC numbers between 8938628 and 10613718.  The red curves are the total model $O-C$ values.  Dark blue is the model fit for the Roemer delay (eq.~(\ref{eqn:Roem})).  Light green curves represent the total physical delay (sum of eqs.~(\ref{eqn:phys1}) and (\ref{eqn:phys2})).  Note that the vertical scales are different on all of the plots; the amplitudes of the $O-C$ curves range from a low of 100 sec to a high of 2000 sec.  The linear and quadratic terms in the fit have been subtracted before the plot is made.}
\label{fig:OmC4}
\end{figure*}

\subsubsection{Physical delay}

The second major effect that results in the eclipse timing variations is the so-called ``physical delay''.  This results from physical changes to the clock, i.e., actual variations in the binary period, caused by the third body.  Qualitatively, the presence of the third body causes the orbital period of the binary to be {\em longer} than it would be in isolation.  The perturbed binary period depends on the instantaneous distance from the center of mass of the binary to the third star, $r_{\rm trip}$, and is longest when $r_{\rm trip}$ is smallest. If the third star is in a circular coplanar orbit, the instantaneous distance $r_{\rm trip}$ is a constant, and there are no first order effects to be observed in the eclipse times since the lengthened binary period is then a constant as well (here we are still assuming a circular inner binary orbit).  However, if the orbit of the third star is either eccentric or inclined with respect to the orbital plane of the binary, then the distance between it and the binary CM and/or the tidal interaction is constantly changing, and so is the binary orbital period.  This leads to a very distinctive $O-C$ curve.

\begin{figure*}
\centering
\includegraphics[width=0.95\linewidth]{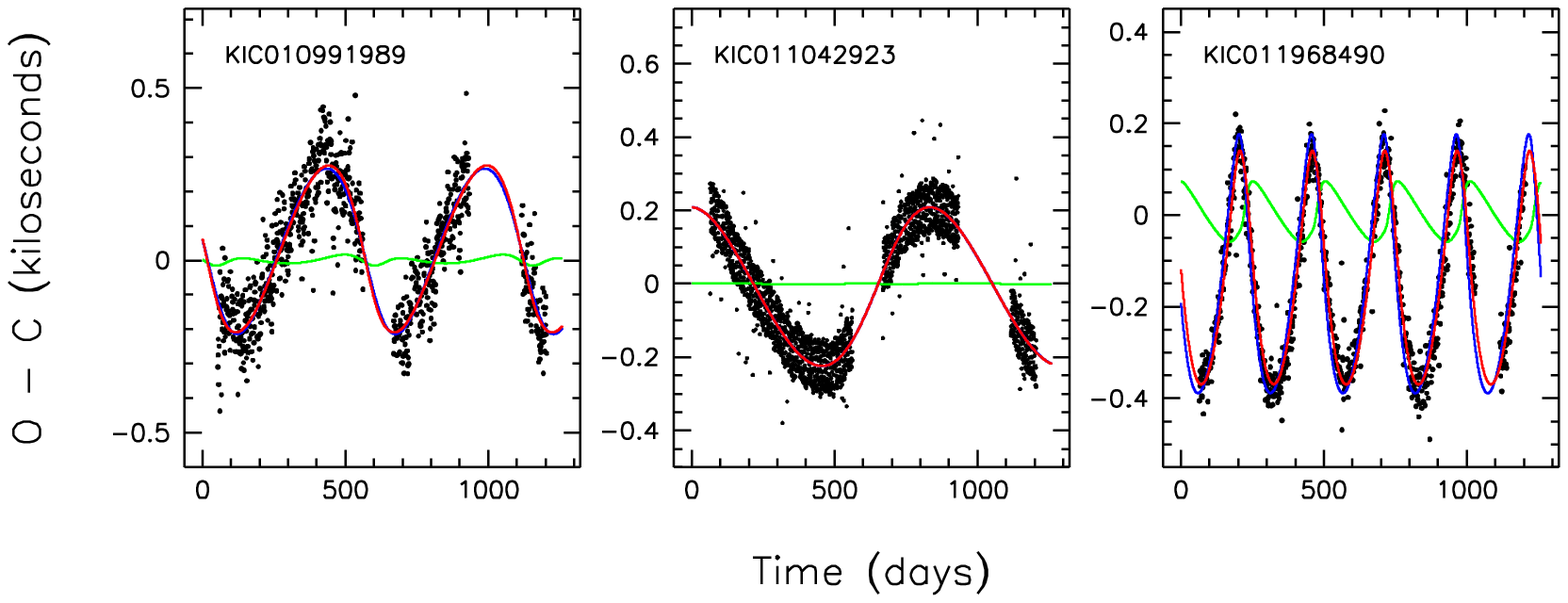}
\caption{$O-C$ data and model fits for 3 systems with KIC numbers between 10991989 and 11968490.  The red curves are the total model $O-C$ values.  Dark blue is the model fit for the Roemer delay (eq.~(\ref{eqn:Roem})).  Light green curves represent the total physical delay (sum of eqs.~(\ref{eqn:phys1}) and (\ref{eqn:phys2})).  Note that the vertical scales are different on all of the plots; the amplitudes of the $O-C$ curves range from a low of 200 sec to a high of 300 sec.  The linear and quadratic terms in the fit have been subtracted before the plot is made.}
\label{fig:OmC5}
\end{figure*}

A number of approximate analytic expressions have been developed for the case of a third body perturbing the orbit of a circular binary (see, e.g., Brown 1936; Harrington 1968; 1969; S\"oderhjelm 1975, 1982, 1984; Borkovits et al.~2003; Agol et al.~2005; Borkovits et al.~2011) on the timescale of the orbital period of the triple.  The perturbative calculation takes advantage of the hierarchical nature of the system and expands the equations of motion in terms of the small parameter $\xi = r_{\rm bin}/r_{\rm trip}$, where $r_{\rm bin}$ is the instantaneous separation of the two stars in the binary and $r_{\rm trip}$ is the instantaneous distance from the tertiary star to the CM of the binary, as defined above. The short period perturbations (those on the timescale of the binary period) are of small amplitude (higher order in $\xi$) and less interesting observationally; averaging over the binary period results in an expression for the slower (but higher amplitude) variations in the perturbed period of the binary on the timescale of $P_{\rm trip}$.

\begin{figure}
\begin{center}
\includegraphics[width=0.99\columnwidth]{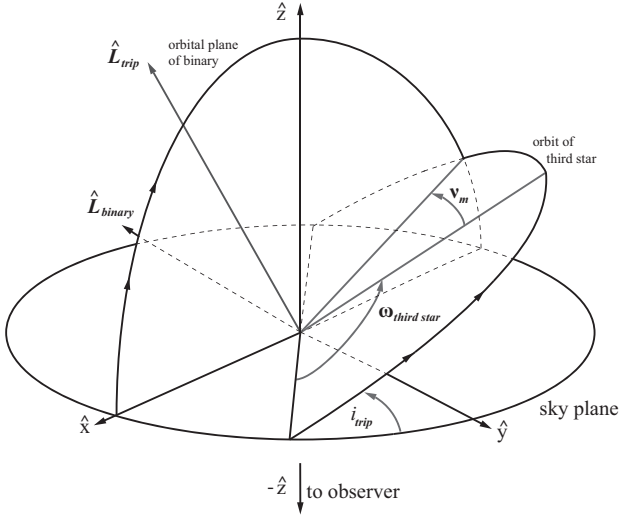}
\caption{Geometry of the triple system.  The observer is viewing along the $+\hat{z}$ axis, and the $xy$ plane coincides with the plane of the sky.  For the purpose of this diagram, as well as for our analysis, we take the binary orbit to be circular and its orbital angular momentum vector to lie approximately in the $xy$ plane.  Of the four angles used in the analysis, $i_{\rm trip}$, $\omega$, $v_m$, and $i_m$, the first three are indicated in the diagram, while $\cos i_m \equiv \hat{L}_{\rm bin} \cdot \hat{L}_{\rm trip}$.  (Note, however, $\omega \equiv \omega_{\rm third~star}+\pi$.)  In words, $i_{\rm trip}$ is the conventional inclination angle of the orbital plane of the third-star; the mutual inclination angle, $i_m$, is the angle between the two orbital planes; $\omega$ is the angle along the outer orbit of the binary CM from the plane of the sky to the periastron point; and $v_m$ is the angle along the outer orbit from periastron of the third star in its orbit to the plane  of the binary. }
\label{fig:geometry}
\end{center}
\end{figure}

The most comprehensive of the expressions for the physical delay in the case of circular binaries\footnote{In this work we utilize two pieces of information to constrain the orbital eccentricity of the binaries within our candidate triple stars: (i) analysis of the epoch-folded light curves (see Table 1 and \S \ref{sec:binaryLC}); (ii) the similarity of the $O-C$ curves for the primary and secondary eclipses for the vast majority of the systems (especially those with $P_{\rm bin} \lesssim 2$ days) provides additional evidence for the approximate circularity of the binary orbits (see Table 1).} is given in Borkovits et al.~(2003; but see also Borkovits et al.~2011 for a more expansive treatment of perturbations to eccentric binaries). The expression there encompasses the perturbations to the period of the binary occurring on a timescale equal to $P_{\rm trip}$, and consists of three terms, of which we use two. The two terms appearing in the $O-C$ formula which we use are:
\begin{eqnarray}
\frac{P_1(t)}{A_{\rm phys}}  =  \left(2\,\mathcal{I}-\frac{2}{3}\right)\left[\phi(t) +e \sin \phi(t) - \theta(t)\right] 
\label{eqn:phys1}
\end{eqnarray}
\begin{eqnarray}
\frac{P_2(t)}{A_{\rm phys}} & = &  \left(1-\mathcal{I}\right) \{\sin \left[2\phi(t)-2v_m\right] \nonumber \\
& +& e \sin \left[\phi(t)-2v_m\right] + \frac{e}{3} \sin \left[3\phi(t)-2v_m\right] \}
\label{eqn:phys2}
\end{eqnarray}
where 
\begin{eqnarray}
A_{\rm phys} = \frac{3}{8 \pi} \frac{M_3}{M_{\rm trip}}\frac{P_{\rm bin}^2}{P_{\rm trip}}\left(1-e^2\right)^{-3/2}
\label{eqn:physamp}
\end{eqnarray}
with the following definitions: $\phi$ and $\theta$ are the true and mean anomalies of the orbit of the triple star system, $\mathcal{I}$ is $\cos^2 i_m$ with $i_m$ the mutual inclination of the binary orbital plane with respect to the orbital plane of the triple, and $v_m$ describes the orientation of the periapse of the triple star system with respect to the binary plane. (See Fig.\,\ref{fig:geometry} for definitions of the parameters describing the system geometry.)

The third term in this sequence (not given here), $P_3(t)$, is proportional to $\cot i_{\rm bin} \sin i_m$, where $i_{\rm bin}$ is the inclination to the plane of the sky of the binary orbit. Given that the binaries we are studying exhibit eclipses, $\cot i_{\rm bin}$ is likely to be small.  If, in addition, the mutual inclination angle of the two orbital planes is small, then the product of $\cot i_{\rm bin} \sin i_m$ is likely to be negligible for our purposes.  Thus, in the present work, we exclude this third term.  

\begin{figure}
\begin{center}
\includegraphics[width=0.99\columnwidth]{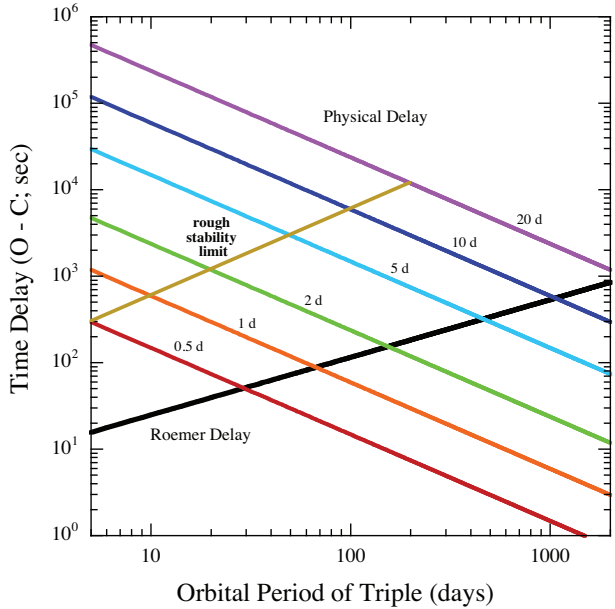}
\caption{Comparison of the Roemer amplitude (black curve), given by eq.~(\ref{eqn:Roemamp}) and the physical amplitude (colored curves) given in eq.~(\ref{eqn:physamp}) as a function of the orbital period of the triple system.  The various physical delay curves are for different assumed binary periods, ranging from 0.5 days to 20 days, as labeled. See text for a list of the nominal values that were assumed for the other parameters in eqs.~(\ref{eqn:Roemamp}) and (\ref{eqn:physamp}). Dynamically stable systems would be expected to lie below and to the right of the gold curve (see eq.~\ref{eqn:stableP}).}
\label{fig:Roem_Phys}
\end{center}
\end{figure}

As an illustration of how the Roemer and physical delays compare, we show in Fig.\,\ref{fig:Roem_Phys} a plot of the amplitudes of the Roemer and physical delays as a function of $P_{\rm trip}$ for six different assumed periods of the binary.  We adopted illustrative values of $e = 0.3$, $i_{\rm trip} = 60^\circ$, and all masses equal to 1 $M_\odot$.  As could be inferred from the analytic expressions, the Roemer delay dominates for longer orbital periods of the triple system and shorter binary periods, and vice versa for the physical delay.  The two effects are roughly comparable for a 1-year period of the triple star system and a binary with a 1-2 day period.  

Finally, we note that the accuracy of these analytic expressions (eqns.~\ref{eqn:phys1} and \ref{eqn:phys2}) has been checked in the original Borkovits et al.~papers (2003, 2007, 2011) via direct 3-body numerical integration.  However, one might expect that these formulae, derived assuming the parameter $\xi = r_{\rm bin}/r_{\rm trip}$ is small, must break down if the pericenter passage of the third star is too close.  In particular, a very close passage of the third star could induce a substantial eccentricity in the binary orbit.  The formulae above, derived assuming a circular binary orbit, would then not apply. We find that, for coplanar orbits, the formulae agree well with numerical experiments as long as:
\begin{eqnarray}
a_{\rm trip} (1-e) \gtrsim 5 \, a_{\rm bin}
\end{eqnarray}
Here $a_{\rm trip}$ and $e$ are the full semimajor axis of the orbit of the triple system and its corresponding eccentricity, and $a_{\rm bin}$ is the orbital separation of the two stars in the binary.  In terms of the orbital periods, this corresponds to 
\begin{eqnarray}
P_{\rm trip} (1-e)^{3/2} \gtrsim 14 \, P_{\rm bin}
\end{eqnarray}
for an assumed set of three equal mass stars.  

An exception to this agreement between the analytic expression and the numerical results can occur when longer-term perturbations (discussed below in \S \ref{sec:longterm}) set in.  Since the timescales for these longer-term perturbations are typically in the range of a decade to centuries (see Table 3),  they can be fitted (or, effectively removed) by simply adding linear and quadratic terms to the fitting parameters (see \S \ref{sec:analysis}).

\subsection{Longer-Term Perturbations}
\label{sec:longterm}

In addition to the perturbations to the orbital period of the binary that are discussed above and have a complete cycle time equal to the orbital period of the triple system, there are other perturbations that occur on typically much longer timescales.  These include precession of the orbital plane of the binary and possible precession of the longitude of periastron of the binary, if the binary is eccentric.  The approximate timescale for these longer-term perturbations is 
\begin{eqnarray}
\tau_{\rm longterm} ~\propto~ \frac{P_{\rm trip}^2}{P_{\rm bin}}~\frac{M_{\rm trip}}{M_3}\,(1-e^2)^{3/2}
\label{eqn:longterm}
\end{eqnarray}
(Harrington 1968; 1969; Mazeh \& Shaham 1979; Ford, Kozinsky, \& Rasio 2000; Borkovits et al.~2003; Borkovits et al.~2007).  Additionally, if the mutual orbital inclination angle satisfies
\begin{eqnarray}
\sin ^2 i_m > 2/5~~~~{\rm or}~~~~ 39.2^\circ \lesssim i_m \lesssim 140.8^\circ
\end{eqnarray}
Kozai cycles (Kozai 1962) may set in.  In this effect there is a cyclic tradeoff between the growth of orbital eccentricity of the binary (including when it initially has $e_{\rm bin} = 0$) and a corresponding decrease in $i_m$.  If the timescale for this cycle, which is the same as $\tau_{\rm longterm}$ in eq.~(\ref{eqn:longterm}), is longer than the timescales that characterize other perturbations that drive precession of the longitude of periastron in the binary, the Kozai cycle will not operate (Eggleton \& Kiseleva-Eggleton 2001; Fabrycky \& Tremaine 2007).  Moreover, effective damping from the two stars in the binary can terminate the Kozai cycles completely -- preferentially leaving $i_m$ in the range of $35^\circ$ to $50^\circ$ (Fabrycky \& Tremaine 2007).

The values of $\tau_{\rm longterm}$ for all of our triple star candidates are listed in Table 3.  They range from $\sim$3 years to 5000 years, but with only 7 of the systems having $\tau_{\rm longterm} < 15$ years.  Therefore, the generally sinusoidal behavior of these long-term perturbations will look approximately linear or quadratic on the 3-year timescale of the {\em Kepler} data set.  And, as a rough approximation for representing such behavior, we have included a quadratic term in our fit (see \S \ref{sec:analysis}).

\section{Analysis Code}
\label{sec:analysis}

\subsection{Choice of Fitting Parameters}

Given the above expressions for the Roemer and physical delays contributing to the $O-C$ curves, there are a total of 11 free parameters to fit for, under the assumption that the binary orbit is circular.  These include 8 parameters which describe the triple system as an {\em equivalent binary} composed of the third star and a star of mass $M_{\rm bin}$ at the location of the center of mass (CM) of the close binary, and 3 other parameters that describe the $O-C$ curve in the absence of the Roemer and physical delays, i.e., a reference time, slope, and curvature terms: 

\vspace{0.2cm}
\noindent
$e$, eccentricity of the orbit of the triple star system \\
$\omega$, longitude of periastron of the binary CM \\
$\tau$, time of periastron passage in the orbit of the triple \\
$i_m$, mutual inclin. of the orbital planes -- eqs.~(\ref{eqn:phys1}), (\ref{eqn:phys2}) \\
$v_m$, orientation parameter -- eqs.~(\ref{eqn:phys1}), (\ref{eqn:phys2}); see Fig.~\ref{fig:geometry} \\
$P_{\rm trip}$, orbital period of the triple \\
$M_3/M_{\rm trip}$, mass ratio $\propto A_{\rm phys}$ (see eq.~\ref{eqn:physamp}) \\
$f(M_3)^{1/3} =$ cube root of mass function $\propto A_{\rm Roem}$   \\
$t_0$, reference time (time of first binary eclipse) \\
$\Delta P_{\rm bin}$, mean slope of $O-C$ curve $\times ~P_{\rm bin}$ \\
$\dot P_{\rm bin}$, quadratic term 

\vspace{0.15cm}
\noindent
We have chosen to fit for the mass ratio and cube root of the mass function since they are the directly measured quantities via the physical and Roemer delays, respectively, if we know the orbital period of the triple. The orbital period can generally be estimated very well before doing the fit by examining the periodicity of the $O-C$ term.  The $t_0$ term is essentially a measure of the time of the first eclipse in the sequence.  $\Delta P_{\rm bin}$, related to the mean slope of the $O-C$ curve, is not generally zero because we used the binary period in the Slawson et al.~(2011) catalog -- based on only 120 days of data -- to compute the initial set of $O-C$ curves.  Finally, the quadratic term could be used to measure the orbital decay or expansion of the binary; however, we do not expect this effect to be detectable over the course of only a few years.  Rather, we use this quadratic term to take into account possible perturbations that occur on timescales substantially longer than $P_{\rm trip}$ (see, \S \ref{sec:longterm}).

Depending on the Roemer and physical amplitudes, certain among the system parameters may be determined much better than others. For example, if the Roemer delay is dominant and the physical delay is negligible, the mass function will be well determined but the parameters $i_m$, $v_m$, and $M_3/M_{\rm bin}$ will {\em not} be substantially constrained. On the other hand, if the physical delay is well measured but the Roemer amplitude is small, then the mass ratio, $M_3/M_{\rm trip}$ will be more tightly constrained, while the mass function and longitude of periastron will be ill defined.  
 
For a number of reasons we decided against using either a conventional Levenberg-Marquardt (LM) or Monte Carlo Markov Chain (MCMC) fitting procedure.  First, we note that there are two different functions (i.e., physical and Roemer delays) possibly contributing to the structure of the $O-C$ curve, and one does not know, a priori, how much each contributes.  Specifically, in most cases, the two functions are not typically orthogonal, and therefore they can trade off against one another in the fit.  As a result, there can be very large regions in parameter space that yield comparably good fits.  Second, given the large number of systems to deal with, we want to search all of parameter space and estimate the uncertainties at the same time.  The LM method is not particularly good for exploring parameter space with highly and nonlinearly structured correlation functions among the parameters.  The MCMC fitting technique is not ideal for exploring wide ranges of parameter space, especially when trying to fit 39 systems.

We therefore constructed a simpler, though less formal, Monte Carlo fitting code that is better suited to the task of fitting 39 systems in an automated, hands-off fashion.  In this approach we choose a random value for each of the following 7 parameters: $e$, $\omega$, $\tau$, $v_m$, $P_{\rm trip}$, $M_3/M_{\rm trip}$, and $f(M_3)^{1/3}$.  The parameters are chosen with a uniform distribution over their entire plausible ranges.  The remaining 4 parameters: $i_m$, $t_0$, $\Delta P_{\rm bin}$, and $\dot P_{\rm bin}$ can then be determined via a simple matrix inversion since they appear linearly in the fitting function.  (Actually, in the case of $i_m$, it is $\cos^2 i_m$ that appears linearly in the equations.)

The uncertainty on the individual data points is determined empirically as follows.  All data points for a given system are assumed to be equally weighted.  We then make a first-pass run with our simple MC fitting code to find a good set of system parameters. Using that fit, we scale the size of the error bars so that the normalized value of the chi-squared statistic, $\chi^2_{\nu}$, is equal to 1.  From then on, each time the code is run, we use that same value for the error bars on the individual points (unless subsequent runs find a substantially improved fit).  

In all subsequent runs, the code operates as follows.  If the value of $\chi^2_\nu$ resulting from a particular selection of parameters is $\chi^2_\nu > 1.3$ then we add the ratio of likelihoods, $\exp[-(\chi^2-\chi_0^2)/2]$ (where $\chi_0^2$ is the value for the best fit), to the various probability histograms that are being accumulated for each parameter.  The code then chooses another random set of possible system parameters.  If, on the other hand, the value of $\chi^2_\nu$ resulting from a particular selection of parameters is $\chi^2_\nu < 1.3$, then the code does an additional 1000 draws for a more restricted range of the parameters surrounding the particular choice of parameters that yields the ``good'' $\chi^2$ value.  When the 1000 additional draws have been completed, and the ratio of likelihoods has been recorded for each draw, the broad grid search resumes until another combination of parameters is found that yields a value of $\chi^2_\nu < 1.3$.  At that point, another 1000 localized draws are made, and so forth.  With this prescription, on average, about half the draws cover the broad search while the other half covers a more restricted range of parameters.  

This analysis scheme seems reasonably optimum in terms of covering all of parameter space while exploring in greater detail the regions which yield the best fits.  Without full or rigorous justification, we also expect it to give approximately correct estimates of the parameter uncertainties.

\subsection{The Fitting Runs}

The number of eclipse times, over 13 {\em Kepler} quarters, to be analyzed in any given binary ranges from only $\sim$40 to as many as 2400, depending on the orbital period (except for the special case of KIC 10319590 where there are only 19 primary eclipses; see Fig.\,\ref{fig:10319}).  The analysis time is essentially linearly proportional to the number of eclipses.  We chose to have the code spend roughly the same amount of {\em time} analyzing each source rather than drawing the same number of random sets of parameters to test.  The reason is that for the shorter binary periods, the $O-C$ curves become dominated by the Roemer delay (since $A_{\rm phys} \propto P_{\rm bin}^2$ whereas $A_{\rm Roem}$ is independent of $P_{\rm bin}$).  Since the Roemer delay has one fewer free parameter, and is generally simpler in shape than the physical delays, such $O-C$ curves can be fit more quickly.  

With this in mind, we typically draw $10^7$ random sets of parameters for a fiducial 5-day binary, and this number is scaled proportionally to $P_{\rm bin}$ from that value.  The analysis then takes a day and a half on a MacBook Air computer, for the full set of 39 systems, and is adequate to yield good fits and system parameters with their uncertainties.  The same analysis was done using $10^6$, $10^7$, and $10^8$ draws (scaled to $P_{\rm bin}/5$ days).  We found that the $10^7$ and $10^8$ draw runs resulted in the substantially the same best fit parameter estimates and any deviations were almost always within the 10\%$-$90\% uncertainty interval.  

\subsection{Test of the Code}

In order to check the basics of the code we simulated eclipse timing data for a number of different triple star systems using a 3-body numerical integrator.  These include cases where the Roemer delay dominated, where the physical delay dominated, and where the two effects were comparable.  White noise of rms amplitude equal to 60 sec was added to the simulated eclipse arrival times.  The artificial data were then analyzed in exactly the same way as the actual $O-C$ data.  The results were that the fitting code recovered the correct input parameters from the simulation, to within the 10\% -- 90\% error constraints (the same as we list in Tables 2 and 3).

\section{Results}
\label{sec:results}

\subsection{Overview}

The results of the automated fits to the 39 triple star candidates are shown in five multi-panel figures (Figs.~\ref{fig:OmC1}-\ref{fig:OmC5}).  They are arranged simply in order of their KIC number.  In each panel, the red curve is the overall fit to the $O-C$ curve, and is the sum of the Roemer and physical delays, which are shown separately as the blue and green curves, respectively.

The fitted parameters and their uncertainties are listed in Tables 2 and 3 along with the 10\% and 90\% (lower and upper) confidence limits.  Table 2 gives, in addition to the binary period, four quantities related to the masses which are derived entirely from fitting the $O-C$ curves.  These are the mass ratio, $M_3/M_{\rm trip}$, the mass function, $M_3^3 \sin^3 i/(M_3+M_{\rm bin})^2$, and the quantities $M_3 \sin^3 i$ and $M_{\rm bin} \sin^3 i$, derived from the mass ratio and mass function -- to the extent allowed by the uncertainties.  We also list the amplitudes of the Roemer and physical delays (the 10\% and 90\% probability limits are given in curly brackets).

In Table 3 the remainder of the fitted parameters, eccentricity, $e$, and time of periastron passage, $\tau$ (relevant to both Roemer and physical delays), the longitude of periastron, $\omega$ (appearing in the Roemer delay only), and the mutual orbital inclination angle, $i_m$, and orientation angle, $v_m$ (both related to the description of the physical delay), are given.  Table 3 also lists the rms of the residuals with respect to the best fitting $O-C$ curve, as well as the calculated timescale for longer-term perturbations (see eq.~\ref{eqn:longterm}).  

A perusal of Figs.~\ref{fig:OmC1}-\ref{fig:OmC5} as well as Table 2 shows that 19 of the $O-C$ curves are dominated by the Roemer delay, 11 are dominated by the physical delay, while the remaining 9 objects have more competitive Roemer and physical amplitudes (here ``dominant'' is defined as a $\gtrsim 3:1$ ratio).  If ``dominant'' is defined by a ratio of $\gtrsim 5:1$, then the corresponding numbers are 18 Roemer, 8 physical, and 13 comparable.  The Roemer delay dominated systems all have binary periods of $\lesssim 2$ days, consistent with the diagram in Fig.~\ref{fig:Roem_Phys}.  Conversely, all the systems with the longer orbital periods (e.g., $\gtrsim 5$ days) are dominated by physical delays.

\subsection{System Parameter Constraints}
\label{sec:parms}

A review of Table 2 will show that for systems that are dominated by the Roemer delay, the cube root of the mass function is indeed determined with  greater fractional accuracy ($\sim$10\%) than is the mass ratio (typically $\gtrsim$40\%).  This follows from the fact that the Roemer amplitude is directly proportional to the cube root of the mass function.  Additionally, in this circumstance, the parameters $\omega$, $\tau$, and $e$ are all relatively well determined, but the parameters strictly associated with the physical delay, $v_m$ and $i_m$, are generally poorly constrained.  Conversely, for the systems where the physical delay dominates, the mass ratio, $M_3/M_{\rm trip}$, is determined to a substantially better fractional accuracy ($\sim$30\%) than is the cube root of the mass function (typically $\gtrsim$50\%).  Again, this is due to the fact that the physical amplitude is directly proportional to the mass ratio.  As well, the parameters $i_m$, $\tau$, and $e$, are better determined than $\omega$ which is only relevant to the Roemer delay.  The parameter $v_m$, generally seems not well constrained, except in six systems -- all ones with dominating physical delays.

\begin{figure}
\begin{center}
\includegraphics[width=0.99\columnwidth]{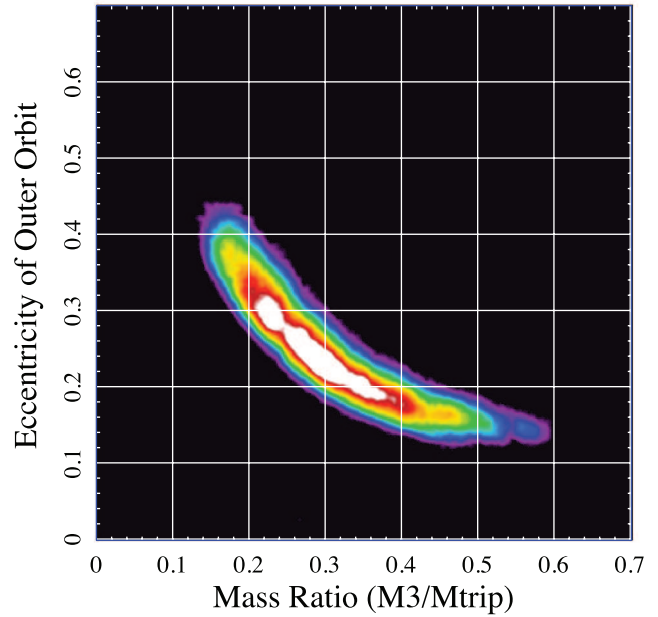}
\caption{Example of the correlation between the eccentricity of the orbit of the triple star system, i.e., the outer orbit, and the mass ratio, $M_3/M_{\rm trip}$ for a system in which the physical delay dominates: KIC 9714358. The colors are scaled according to the relative probability with white and red the highest, blue and purple the lowest.}
\label{fig:corr_e_mr}
\end{center}
\end{figure}

\begin{figure}
\begin{center}
\includegraphics[width=0.99\columnwidth]{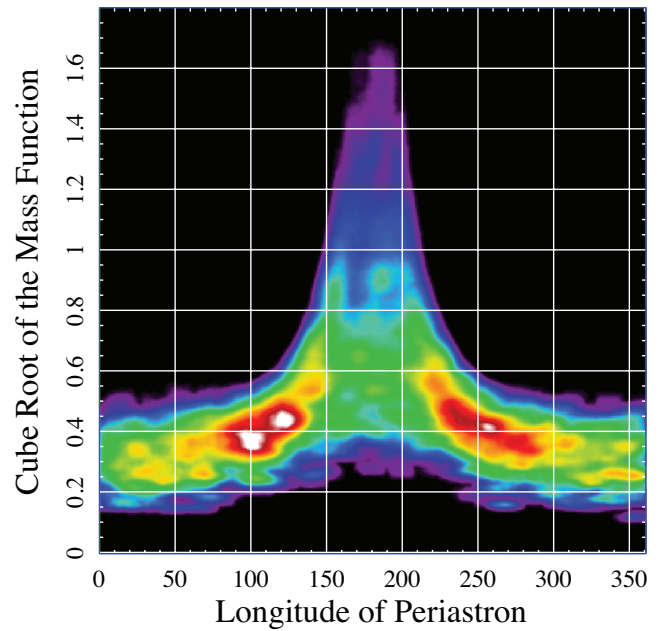}
\caption{Example of the correlation between the cube root of the mass function, $f(M_{\rm 3})^{1/3}$, and the longitude of periastron, $\omega$, of the orbit of the triple for a system in which the physical and Roemer delays are comparable: KIC 9451096. The colors are scaled according to the relative probability with white and red the highest, blue and purple the lowest.}
\label{fig:corr_fm_w}
\end{center}
\end{figure}

One might guess that for those 9-13 systems where the Roemer and physical delays are more comparable (smaller than 3:1 or 5:1 ratios, respectively) both the mass ratio and mass function could be well determined.  This does {\em not} appear to be the case in practice.  The reason is due to the fact that the two sets of functions representing these delays are not substantially orthogonal, and therefore the two functions can add in different ways, consistent with the constraints on the parameters $\tau$, $\omega$, and $v_m$ to produce the total observed amplitude.  It turns out that the Roemer and physical delays, when comparable, can vary {\em together} in amplitude over a fairly wide range while the longitude of periastron, $\omega$, in turn, changes their relative phase in such a way that the sum of the two functions adds to be roughly a constant (and thereby matches the observed $O-C$ curve; see \S \ref{sec:correl} for details).  Thus, in no specific system do we obtain very tight constraints on both $M_3 \sin^3 i$ and $M_{\rm bin} \sin^3 i$ (i.e., with both being determined to better than, e.g., 20\%).

When either the Roemer or physical delay dominates, this type of correlated behavior may be may be present but is much less pronounced (see \S \ref{sec:correl}).  The reason is that for given binary and triple system periods, as well as eccentricity, the physical delay has a tight upper limit that is proportional to $M_3/M_{\rm trip}$ which, by definition, can never exceed unity.  Since the physical delay amplitude is proportional to $P_{\rm bin}^2$, while the Roemer amplitude is independent of $P_{\rm bin}$, for short period binaries (i.e., $\lesssim 0.7$ days) it becomes difficult for the physical amplitude to contribute much to the $O-C$ curves, notwithstanding any issues of orthogonality.  Conversely, the Roemer delay is proportional to the cube root of the mass function which is limited to be less than $M_3$.  While in principle, it is possible for the mass of the third star to take on any value, unless it is a fairly evolved giant, it is unlikely to have a mass greater than a few $M_\odot$ since both stellar radius and $T_{\rm eff}$ were constrained by the nature of the stars selected for inclusion in the KIC (Batalha et al.~2010).  Therefore, for systems with long binary orbital periods the magnitude of the Roemer delay will generally be much smaller than that of the physical delay, even if the shape of the $O-C$ curve matches the expected shape of the Roemer delay.  

\subsection{Correlations Among the Parameters}
\label{sec:correl}

We have tried to select a convenient, consistent set of parameters to fit for all of our candidate triple star systems, regardless of whether they are dominated by the Roemer or physical delays.  It is somewhat inevitable that some of the parameters can become substantially correlated (see discussion in \S \ref{sec:parms}) when the physical delay dominates, vice versa, or even when the two effects are comparable.  Here we show two examples of this type of correlation taken from our Monte Carlo fitting code.  In Fig.~\ref{fig:corr_e_mr} we show the correlation between the eccentricity of the orbit of the triple system (i.e., the outer orbit) and the mass ratio, $M_3/M_{\rm trip}$, for the example of KIC 9714358 which is dominated by the physical delay.  In the case of physical delay only, the amplitude is roughly proportional to the product of these two quantities, and we then expect just such a correlation as is seen in Fig.~\ref{fig:corr_e_mr}.  This can be shown analytically for the case of coplanar orbits from eq.~(\ref{eqn:phys1}) where the term in square brackets on the right hand side, $\left[\phi(t) +e \sin \phi(t) - \theta(t)\right] $, can be expanded in a series for small eccentricities as $\sim$$3e \sin \phi(t)$ (Murray \& Dermott 2000), while the $M_3/M_{\rm trip}$ part of the proportionality is found in eq.~(\ref{eqn:physamp}). For non-coplanar orbits, one of the terms in eq.~(\ref{eqn:phys2}) is not proportional to $e$ while the other two terms are; therefore, the correlation becomes less pronounced as the mutual orbital inclination increases.

We now consider the key correlation for the case where the physical and Roemer delays are more comparable.  In Fig.~\ref{fig:corr_fm_w} we show the correlation between the cube root of the mass function, $f(M_3)^{1/3}$ and the longitude of periastron of the outer orbit, $\omega$, for the case of KIC 9451096.  The correlation seen in Fig.~\ref{fig:corr_fm_w} is quite strong and symmetric around 180$^\circ$.  The zero delay point of the physical delay typically occurs near the time of periastron passage, $\tau$ (especially as $i_m \rightarrow 0$), while the Roemer delay is zero at $\sim \tau-\omega P_{\rm trip}/2 \pi$.  Therefore, if $A_{\rm Roem} \simeq A_{\rm phys}$ the two functions will have a combined amplitude $A_{O-C} \simeq 2 A_{\rm Roem} |\cos(\omega/2)|$.  It then follows that $A_{\rm Roem} \simeq A_{\rm phys} \simeq \frac{1}{2} A_{O-C}/ |\cos(\omega/2)|$ and these two parameters ($A_{\rm Roem}$ and $\omega$) are thus highly correlated, as seen in Fig.~\ref{fig:corr_fm_w}.

\begin{figure}
\begin{center}
\includegraphics[width=0.99\columnwidth]{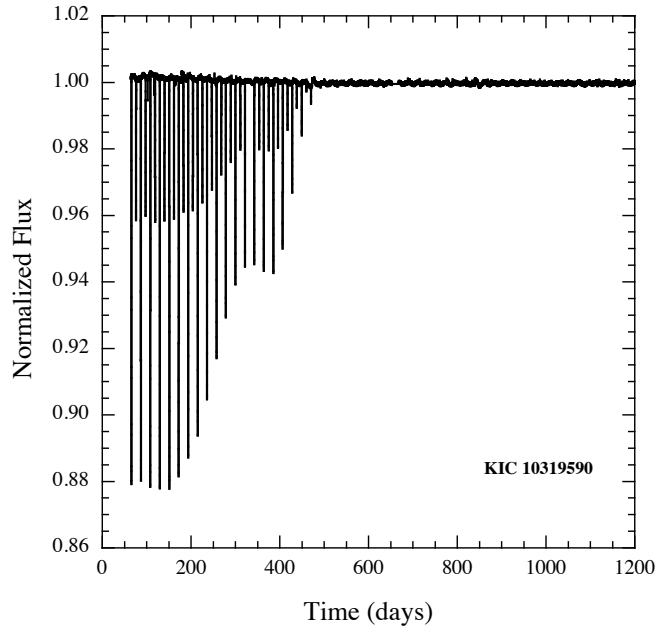}
\caption{Example of a system (KIC 10319590) where the eclipse depths exhibit strong variations with time.  In this extreme case, the eclipses completely disappear after $\sim$400 days, presumably due to the precession of the binary orbital plane caused by the presence of the inferred third body.  }
\label{fig:10319}
\end{center}
\end{figure}

\subsection{Dynamical Stability of Orbits}

We mention in passing that, as a sanity check on the orbital solutions we have found, the mutual orbits of the three stars would be expected to have long-term dynamical stability.  The stability criteria for triple systems have been studied for  decades, and are conveniently summarized by Mikkola (2008).  In particular, we cite here the expression due to Mardling \& Aarseth (2001):
\begin{eqnarray}
a_{\rm trip}  \gtrsim 2.8 \left(\frac{M_{\rm trip}}{M_{\rm bin}}\right)^{2/5}\frac{(1+e)^{2/5}}{(1-e)^{6/5}} ~ a_{\rm bin}
\label{eqn:stableA}
\end{eqnarray}
where, again, $e$ is the eccentricity of the orbit of the triple system.  Expressed in terms of the orbital periods, this stability criterion comes to:
\begin{eqnarray}
P_{\rm trip}  \gtrsim 4.7 \left(\frac{M_{\rm trip}}{M_{\rm bin}}\right)^{1/10} \frac{(1+e)^{3/5}}{(1-e)^{9/5}} ~ P_{\rm bin}
\label{eqn:stableP}
\end{eqnarray}  
Note that, while we do not know the masses of the binary and triple very accurately, the dependence on masses in eq.~(\ref{eqn:stableP}) is extremely weak.  Moreover, in most cases we have a good handle on $e$, and an excellent measurement of both $P_{\rm bin}$ and $P_{\rm trip}$.  Direct computation then shows that all of our triple star candidates are nominally stable. This is another sanity check that suggests that these are true triple stars and not false positives, since false positives should not be biased towards satisfying stability requirements.

\subsection{Supplemental Information Required}

\begin{figure}
\begin{center}
\includegraphics[width=0.99\columnwidth]{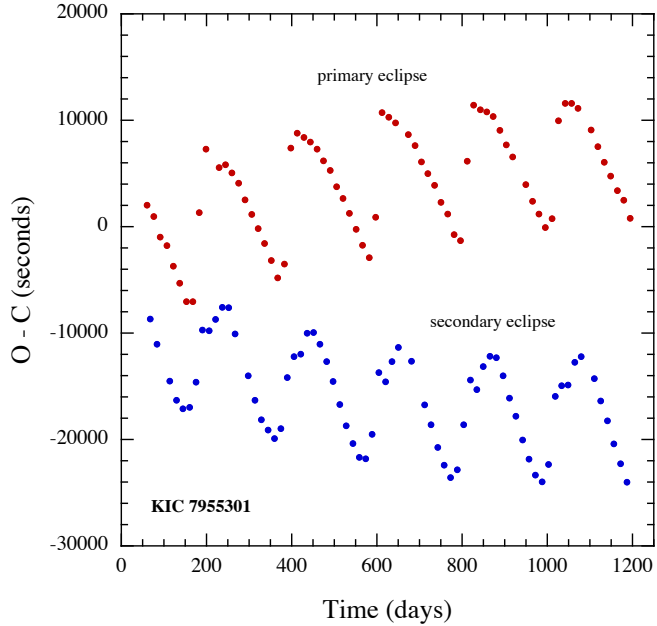}
\caption{Example of a system (KIC 7955301) where the $O-C$ curves for the primary and secondary eclipses lie on ``divergent'' paths -- at least for the 3-year duration of the Q1-Q13 data. As well, the two $O-C$ curves even have somewhat different profiles. }
\label{fig:diverge}
\end{center}
\end{figure}

Supplemental information will be required in order to reasonably infer full sets of system parameters with astrophysically useful accuracy for the triple star candidates identified in this work.  For some of the systems there can be up to three pieces of supplemental information from the {\em Kepler} light curves themselves.  It is beyond the scope of this paper to try to utilize this information, but we list them here for the interested reader.  Seven of the systems exhibit secularly varying eclipse depths (see Table 1).  The most extreme case of secularly varying eclipse depths is the case of KIC 10319590 whose flux vs.~time is shown in Fig.~\ref{fig:10319} where the eclipses disappear after $\sim$400 days. Two of the systems show eclipses of, and/or by, the third body (Carter et al.~2013).  Finally, at least five of the systems have $O-C$ curves for the primary and secondary eclipses that are different in shape and/or systematically diverge in phase with respect to one another.  A good example of this latter effect is exhibited in Fig.\,\ref{fig:diverge} for the case of KIC 7955301 where the $O-C$ curves for both the primary and secondary eclipses are shown.  In total, seven systems of the 39 exhibit one or more of these three different features. (See Table 1 for a summary.)

In these seven cases, the supplementary information from the {\em Kepler} photometry can be modeled with a 3-body code to gain a much more complete understanding of the system parameters (see, e.g., Carter et al.~2011 and 2013).

For these seven systems, as well as the remainder of the 39 triple star candidates, it will be important to obtain radial velocity measurements.  Even a high-quality, single-epoch spectrum, could provide significant insight into the nature of the three constituent stars.  Measuring the radial velocities within the binary, and, even better, of all three stars, would lock in most of the physically important system parameters that are only loosely constrained through the eclipse timing analysis alone.  

In general, the binary orbital periods are quite short (only seven have $P_{\rm bin} \gtrsim$ one week), so it will not take a long interval to unravel the properties of the binary (e.g., its masses and luminosity contribution to the triple system).  The orbital periods of most of the triple systems range from 48 days to 1 year.  The median period is $\sim$330 days.  Therefore, radial velocity measurements aimed at determining the properties of the orbit of the triple system would have to span a good portion of the observing season for the {\em Kepler} field.  

\subsection{Binary System Light Curves}
\label{sec:binaryLC}

To gain some further insight into the constituent stars in the 39 systems we have identified, we have constructed folded light curves for each of the binary stars in these systems.  We then used the {\em Phoebe} binary light curve modeling code (Pr\v{s}a \& Zwitter 2005) to fit the binary system parameters, allowing for the ``third light'' parameter (presumably largely due to the light contribution of the third star) to be a variable.  The results for both the contribution of the ``third light'' and the mass ratio of the two stars in the binary, $q_{\rm bin}$, are listed in Table 1.  In principle, this information can be used in conjunction with the constraints on $M_3$ and $M_{\rm bin}$ found from the analysis of the $O-C$ curves (see Table 2) to infer the three masses individually, albeit with wide uncertainties.  

We were also able to use the {\em Phoebe} fits to check the orbital eccentricities of the binary systems as reported by Slawson et al.~(2012), and we find reasonable agreement, though with the {\em Phoebe} values of $e_{\rm bin}$ tending to be a bit lower.  The value of $e_{\rm bin}$ is important for the expected form of the physical delay curve; the $O-C$ curves can be noticeably affected when $e_{\rm bin} \gtrsim$0.05 or so.
Table 1 lists the binary eccentricities computed from values given in the Slawson et al.~(2012) catalog, but replaced in four cases with the {\em Phoebe} result (where the former value of $e_{\rm bin}$ was more than 3 times higher than the {\em Phoebe} value).    In all, six of the systems have $e_{\rm bin} \gtrsim 0.075$, and we note that the fitted triple star parameter values for these could be significantly different from the true system parameters.

\section{Discussion}
\label{sec:discuss}

In all, we computed and examined the $O-C$ curves for some 2000 {\em Kepler} binaries.  We found that approximately 50\% of these yielded quite useful portraits of the source eclipse timing behavior, with typical rms scatter less than 100 seconds.  Some 20\% were contact (or otherwise short-period) binaries that tended to exhibit erratic, or random-walk like behavior that made it difficult to search for periodic signatures of third bodies.  The remaining 30\% yielded at most minimally useful information.  In some cases this latter category could be attributed to eclipse depths that were too small, stellar noise (i.e., starspots, stellar oscillations, etc.) that was not sufficiently filtered out, and/or inadequacies in our eclipse detection algorithm\footnote{The fraction of systems ($\sim$30\%) that yielded no useful $O-C$ curves did not improve with the use of our newly developed, more formal cross correlation analysis (mentioned earlier in the text.)}.  We believe that the 50\% of binaries for which we were able to obtain good eclipse timing information is sufficient so that our findings are not substantially biased.

\begin{figure}
\begin{center}
\includegraphics[width=0.99\columnwidth]{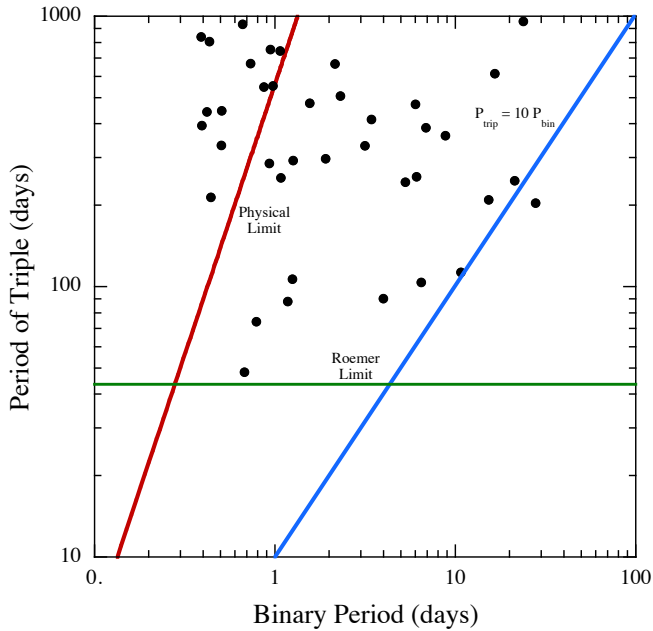}
\caption{Plot of the orbital periods of the candidate triple systems vs.~the period of the binary system they contain.  The blue line indicates the locus of points where $P_{\rm trip}/P_{\rm bin} = 10$, as a representative stability criterion.  Most systems should lie to the left of this line which is taken from eq.~(\ref{eqn:stableP}) with $e = 0.3$.  The horizontal green line is a rough lower limit to values of $P_{\rm trip}$ that can be detected via the Roemer delay with the {\rm Kepler} Q1-Q13 data set, given a sensitivity of $\sim$50 seconds in detectable amplitude (see eq.~\ref{eqn:Roemamp}).   Finally, the red line is a rough upper limit to values of $P_{\rm trip}$ that can be detected via the physical delay with the {\rm Kepler} Q1-Q13 data set given a sensitivity of $\sim$50 seconds in detectable amplitude (see eq.~\ref{eqn:physamp}). An assumed value of $e=0.3$ was used to evaluate this latter limit.  Systems to the left of the red line are typically detected via the Roemer delay.}
\label{fig:periods}
\end{center}
\end{figure}

Notwithstanding the above general statements about our search, there are quite a few observational selection effects in play.  These include the construction of the {\em Kepler} input catalog itself which selected for certain spectral types and radii.  Then, there is the binary detection efficiency for the various stars within the KIC.  Among other things, this depends on stellar pulsations and starspot activity.  Within our search for triples, the depth of the binary eclipses, which in part depends on the brightness of the third star, affects the timing accuracy.  The erratic timing behavior of many contact binaries (at the $\sim$300 sec rms level) makes it harder to detect tertiary companions (via eclipse timing variations) in these systems. Finally, if we limit ourselves to seeing 1.5 -- 2 orbital cycles of the triple system, then orbital periods greater than $\sim$900 days are nearly ruled out.  In fact, in our visual inspection of the set of $O-C$ curves we see numerous such potential longer-period triple star candidates (see also Gies et al.~2012).  On the short period end, there are many beat periods, between the {\em Kepler} cadence and the binary period, up to $\sim$20 or 30 days.  Thus, it is difficult to identify likely real triple star candidates in this period range.    

The periods of the triple star candidates we found are plotted vs. their binary periods in Fig.\,\ref{fig:periods}.  We show a rough dynamical stability bound on the right (blue curve).  This limit is derived from eq.~(\ref{eqn:stableP}) for an assumed typical orbital eccentricity of the triple system equal to 0.3.  Most of the triples should lie to the left of this curve.  If we assume a typical sensitivity in the $O-C$ curves of $\sim$50 sec, the corresponding orbital period of the triple system required to produce a detectable signal purely via the Roemer delay is about 45 days (see green curve in Fig.\,\ref{fig:periods}, above which we should be able to detect the light-travel-time effects).  Here we have assumed all 1 $M_\odot$ constituent stars, and an orbital inclination of the triple system equal to 60$^\circ$ (see eq.~\ref{eqn:Roemamp}).  The limiting triple-star periods for the physical delay are indicated crudely by the red curve in Fig.\,\ref{fig:periods}).  This is based on eq.~(\ref{eqn:physamp}) with $e=0.3$ and all equal constituent masses.  Systems detected via the physical delay should lie to the right of this curve for an amplitude sensitivity in the $O-C$ curves of $\sim$50 sec; systems to left of this line are detected via the Roemer delay.  Finally, it is difficult at best to confirm any triples with $P_{\rm trip} \gtrsim 1000$ days (see also Gies et al.~2012).  

Thus, Fig.\,\ref{fig:periods} indicates that most of the 39 triple star candidates are reasonably well dispersed (in log space) around the zone of detectability and stability.  

Because of the various observational and analysis selection effects alluded to above, it is difficult for us to draw far-reaching conclusions about the fraction of binary systems with relatively close tertiary companions.  However, there are some things we can say in this regard.  Approximately 1000 of the {\em Kepler} binaries yielded useful constraints on the eclipse timing via our particular approach to the analysis.  There were some 39 triple star candidates found among these with $48 \lesssim P_{\rm trip} \lesssim 900$ days, spread roughly uniformly with respect to $\log P_{\rm trip}$.  Without trying to be too precise, we can say that we see evidence for roughly a comparable number of potential candidate triple systems with $P_{\rm trip}$ in the range of $\sim$$1000-2500$ days, where only at most one to a fraction of an orbital cycle is revealed.  This would suggest that perhaps $\sim$8\% of close binaries have tertiary companions that have orbital periods of less than $\sim$7 years.  Again, the $O-C$ sensitivity limit here is $\sim$50 sec (rms scatter) with which we are able to time the eclipses.  

Finally, in terms of the completeness of our initial survey for triple systems, we note that some of the companions to binaries with $P_{\rm bin} \lesssim 1$ day and $P_{\rm trip} \lesssim 30$ days can produce delays that are too small (i.e., less than a few tens of seconds) to be detectable with the current approach.  In particular, note the unpopulated region in the bottom lower left corner of Fig.\,\ref{fig:periods}. 

\begin{figure}
\begin{center}
\includegraphics[width=0.99\columnwidth]{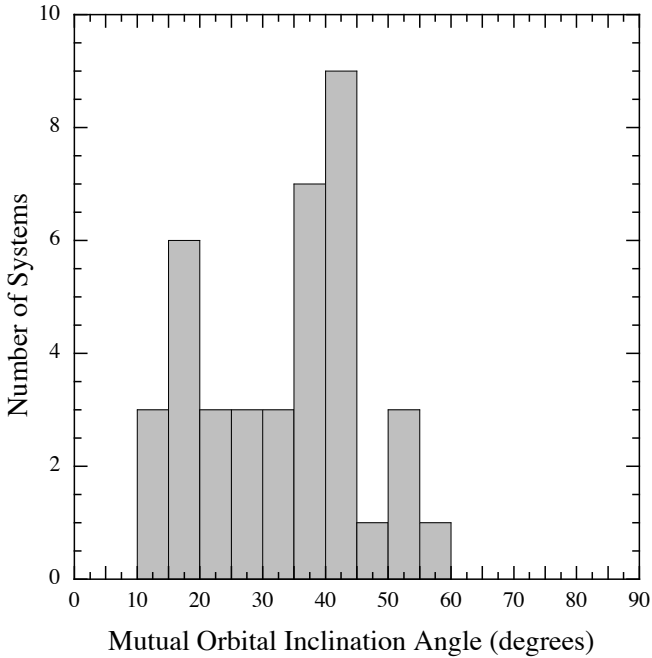}
\caption{Distribution of the mutual inclination angles of the 39 triple star candidates. Note that, in general, the uncertainties in $i_m$ are larger than the $5^\circ$ bin size used for the histogram.}
\label{fig:inclin}
\end{center}
\end{figure}

Among the most popular formation theories for very close binaries (e.g., with $P_{\rm bin} \lesssim 3$ days) are those which invoke a third star, even if quite distant (with $P_{\rm trip}$ up to $10^5$ yr), to effect the closeness of short-period binaries.  These scenarios typically involve so called ``KCTF'' (Kozai cycles with tidal friction;  Eggleton \& Kiseleva-Eggleton 2001; Fabrycky \& Tremaine 2007; but it is also possible that magnetic braking plays a role, e.g., Verbunt \& Zwaan 1981; Matt \& Pudritz 2005).  Fig.~\ref{fig:inclin} shows the distribution of mutual orbital inclination angles in our sample of triple star candidates.  This distribution was produced without regard for the large uncertainties in the measurements of $i_m$ which typically exceed the bin width of 5$^\circ$ used here.  Nonetheless, there is something of a very suggestive peak in the mutual orbital inclination range of $35^\circ - 45^\circ$ predicted by Fabrycky \& Tremaine (2007) for the KCTF scenario.  Within our parameter uncertainties, it is quite possible that the Kozai cycle is no longer operative in any of these systems. 

The present study of tertiary stars orbiting short period binaries is quite complementary to those of others (see e.g., Mazeh 1990; Tokovinin et al.~2006; Pribulla \& Rucinski 2006; D'Angelo, van Kerkwijk, \& Rucinski 2006; Rucinski, Pribulla, \& van Kerkwijk 2007; Raghavan et al.~2010).  In particular, the region of orbital period space covered by Tokovinin et al.~(2006; $20 \lesssim P_{\rm trip} \lesssim 10^5$yr and $P_{\rm bin} \lesssim 25$ days) is almost exactly complementary to ours which extends up to $P_{\rm trip} \lesssim 3$ yr and covers the same range of binary periods (see Fig.~12 in Tokovinin et al.~2006).  If we somewhat arbitrarily adopt a distribution of orbital periods for triple systems that is constant per logarithmic interval, then our detection of $\sim$4\% triples over a factor of 20 in $P_{\rm trip}$ (1.3 dex) is consistent with a significant fraction of all close binaries having tertiary companions (Tokovinin et al.~2006; Pribulla \& Rucinski 2006; Raghavan et al.~2010).  If we assume that possible triple-star periods cover $\sim$20 days $- ~10^5$ yr (6.3 dex), then we have examined $\sim$1/5 of this range.  Therefore, we might speculatively extrapolate our results to suggest that $\gtrsim$20\% of close binaries have tertiary companions.  Tokovinin et al.~(2006) find a much higher fraction for binaries with $P_{\rm bin} \lesssim 3$ days, and a more comparable one to our value for $P_{\rm bin} \gtrsim 12$ days.  Thus, given all the uncertainties, our results may not be dissimilar.  However, we do not have the statistics to comment on the tertiary fraction separately for binary periods above and below this transition period of $\sim$10 days (see, in particular, Fig.~14 of Tokovinin et al.~2006).

\section{Summary and Conclusions}
\label{sec:concl}
 
We have analyzed the {\em Kepler} binary data set for eclipse timing variations, with the intention of identifying signatures of the presence of third bodies.  We found some 39 plausible candidates for triple star systems, eight of which had been previously found by the members of the {\em Kepler} team, but only a few of these had been studied in any detail.  Some were found via tertiary eclipses, while others were detected from systematic variations in their $O-C$ curves (in the latter case using typically only $\sim$1/10 of the data in the current study).  We have subjected all of the 39 systems in this study to an analysis which includes possible Roemer delays as well as physical delays.  All the best fits are physically sensible, though revisions may be necessary when Doppler velocity measurements, for example, become available. 

We have shown that at least 8\% of close binaries have tertiary companions with $P_{\rm trip} \lesssim 7$ years.  This is in agreement with other surveys covering tertiaries in much wider orbits over a larger dynamic range in periods.

In order to fully determine the system parameters in the triple system candidates we have found, radial velocity measurements will be required.  This is already being pursued for a number of the systems (see, e.g., Carter et al.~2011; 2013).  Moreover, for those systems which exhibit other effects of the third body, such as tertiary eclipses, varying binary eclipse depths, and/or the effects of binary eccentricity, there is need for analysis with a 3-body dynamics code. We consider our list of triple star candidates something of a starting point for such more extensive studies, both observationally and in modeling.

 We were gratified to find that this exercise has proven a very good way of finding non-eclipsing triples. 
 \vspace{0.2cm}
 
{\em Note added in manuscript}: Since this manuscript was submitted, we have identified another three triple system candidates: KIC 3454864, KIC 5254230, and KIC 7362751.  These have orbital periods for the triple stars of 758, 109, and 549 days, respectively.  Two are Roemer delay dominated systems while KIC 5254230 is strongly dominated by the physical delay.  We have also become aware of the possibility that our triple star candidates KIC 5264818, KIC 5310387, and KIC 8386865 (with high effective temperatures listed in the KIC; see Table 1) may turn out to be pulsating stars rather than binaries.

\acknowledgments
The authors thank Josh Winn for very helpful discussions.  We acknowledge Kathy Tran who participated in some of the eclipse timing analysis, focusing on the behavior of contact binaries.  The authors are grateful to the {\em Kepler} Eclipsing Binary Team for generating the catalog of eclipsing binaries utilized in this work.  Specifically, we thank Andrej Pr\v{s}a for providing us with a list of newly rejected and  ``uncertain'' binaries, a number of which are now considered to be pulsating stars.  JAC acknowledges support for this work that was provided by NASA through Hubble Fellowship grant HF-51267.01-A awarded by the Space Telescope Science Institute, which is operated by the Association of Universities for Research in Astronomy, Inc., for NASA, under contract NAS 5-26555.  KMD acknowledges support from a National Science Foundation Graduate Fellowship.

\newpage

\begin{deluxetable}{lccccccccccc}
\tablewidth{0pt}
\tabletypesize{\scriptsize}
\tablecaption{\label{tab:xray} Candidate Triple-Star Systems Found in the {\em Kepler} Database}
\tablehead{
	\colhead{Source} &
	\colhead{$P_{\rm bin}$} &
	\colhead{$K_p$$^1$}  & 
	\colhead{$T_{\rm eff}$$^1$} &
	\colhead{Prim. Ecl.} &
	\colhead{Sec. Ecl.} &
	\colhead{$e_{\rm bin}$$^3$} &
	\colhead{$q_{\rm bin}$$^4$} &
	\colhead{$L_3/L_{\rm trip}$$^4$} &
	\colhead{Vary. Ecl.} &
	\colhead{Tertiary} &
    	\colhead{Diverg. Prim.}  \\
	\colhead{} & 
	\colhead{(days)} &
	\colhead{(mag.)} &
	\colhead{(K)} & 
      	\colhead{Depth$^2$} &
	\colhead{Depth$^2$} &
	\colhead{} &
	\colhead{} &
	\colhead{} &
	\colhead{Depths$^5$} &
         \colhead{Eclipses$^5$} &
	\colhead{\& Secon. $O-C$$^5$} 
}
\startdata 
3228863$^6$ &  0.730942 & 11.82 & 6561& 0.440 & 0.220 & 0.034 & 1.20(1) & -- & -- & -- & --     \\
4647652 & 1.064820 & 11.81 & 6265 & 0.077 & 0.021 & 0.078 & 0.24(1) & 0.224(4) & -- & -- &  --   \\
4909707 & 2.302370 & 10.69  & NA & 0.043 & 0.018	& 0.073 & 0.075(1) & 0.163(3) & -- &  --	 &  --    \\
4940201 & 8.81659 & 14.98 & 5284 & 0.027 & 0.013 & 0.083 & 0.045(1) & 0.189(1) & -- & -- & --   \\
5039441 & 2.151390 & 12.92 & 5943 & 0.259 & 0.019 & 0.036 & 0.72(1) & 0.018(2) & -- & -- & -- 	  \\
5128972 &  0.505317 & 13.23 & 5776 & 0.094 & 0.047 & -- & 0.53(2) & 0.207(2) & -- & -- & --      \\
5264818 & 1.905052 & 8.86 & 9212 & 0.013 & 0.011 & -- & 1.43(1) & -- & -- & -- & --   \\
5310387 &  0.441669 & 12.68 & 6520 & 0.113 & 0.109 & -- & 0.45(1) & 0.103(3) & -- & -- & --     \\
5376552 &  0.503819 & 12.86 & 6631& 0.206 & 0.204 & -- & 0.59(2) & 0.008(1) & -- & -- &  --      \\
5384802 &  6.08309 & 13.70 & 6433 & 0.020 & 0.020 & 0.072 & 0.42(1) & 0.076(5) & -- & --  & --   \\
5771589 & 10.74007 & 11.81 & 5927 & 0.0011 & 0.0007 & 0.010$^7$ & 0.03(1) & 0.013(1) & yes & -- & yes   \\
6370665 &  0.932316 & 14.00 & 7386 & 0.090 & 0.075 & -- & 0.52(1) & 0.081(32) & -- & -- & --      \\
6525196 &  3.42060 & 10.15 & 5966 & 0.162 & 0.147 & 0.038 & 0.71(1) & 0.024(1) & -- & -- & --     \\
6531485 &  0.676991 &  15.55 & 5587 & 0.021 & 0.017 & 0.048 & 0.032(1) & 0.084(1) & -- & -- & --     \\
6545018 &  3.99146 &  13.75 & 5594 & 0.291 & 0.226 & 0.075 & 0.77(1) & -- & -- & -- & slight    \\
7289157 &  5.26640 & 12.95 & 5922 & 0.062 & 0.006 & 0.064 & 0.10(1) & 0.299(1) & yes & yes & yes    \\
7668648 & 27.8184 &15.32 & 5875 & 0.232 & 0.094 & 0.074 & 0.49(1) &  0.014(2) & yes & yes & yes    \\
7690843 &  0.786259 &  11.08 & 4827 & 0.049 & 0.020 & 0.059 & 0.05(1) & 0.303(1)& -- & -- & --    \\
7837302 & 23.83530 &  13.72 & NA  & 0.026 & none & 0.17 & 0.010(1) & -- & -- & -- & NA   \\
7955301 & 15.3266 & 12.67 & 4821 & 0.016 & 0.01 & 0.20 & 0.23(1) & 0.031(2) & yes & -- & yes   \\
8023317 & 16.57828 & 12.89 &  5625 & 0.034 & 0.002 & 0.057 & 0.15(1) & $<0.001$ & yes & -- & --     \\
8043961 &  1.559210 & 10.74 & 6348 & 0.207 & 0.170 & 0.028 & 0.62(1) & 0.140(1) & -- & -- & --      \\
8192840 &  0.433547 & 13.47 & 6136 & 0.033 & 0.028 & -- & 0.61(1) & 0.279(3) & -- & -- & --      \\
8386865 & 1.25800 &  12.02  & 8510 & 0.005 & 0.005 & 0.59 & 0.053(3) & -- & -- & -- & --     \\
8394040 &  0.302128 & 14.46 & 5697 & 0.042 & 0.034 & -- & 1.15(2) & 0.53(1) & -- & -- & --      \\
8719897 &  3.15142 & 12.39 & 4906 & 0.195 & 0.176 & 0.061 & 0.23(1) & 0.015(3) & -- & -- & --    \\
8904448 &  0.865981 & 13.88 & 7820 & 0.180 & 0.049 & -- & 0.31(1) & 0.065(6) & -- & -- & --     \\
8938628 &  6.86219 & 13.68 & 5602 & 0.050 & 0.034 & 0.062 & 1.42(1) & 0.037(1) & yes & -- & --    \\
9451096 &  1.25039 & 12.64 & NA & 0.233 & 0.087 & 0.063 & 0.46(1) & 0.062(1) & -- & -- & --     \\
9714358 &  6.47418 & 15.00 & 4825 & 0.185 & 0.012 & 0.041$^7$  & 0.36(1) & 0.031(1) & -- & -- & --   \\
9722737 &  0.418528 & 14.93 & 6517 & 0.102 & 0.088 & -- & 0.50(1) & 0.119(4) & -- & -- & --      \\
9912977 &  0.943916 &  13.73  &  NA & 0.292 & $< 0.015$  & 0.01$^7$ & 0.20(1) & -- & -- & -- & -- \\
10095512 &  6.01720 & 13.05 & 5795 & 0.113 & 0.051 & 0.082 & 0.77(1) & 0.030(1) & -- & -- & --    \\
10226388 & 0.660658 & 10.77 & NA & 0.174 & 0.131 & -- & 0.18(1) & -- & -- & -- & --   \\
10319590 & 21.3216 & 13.73  & 5518 & 0.026 & 0.008 & 0.108 & 0.40(1) & 0.079(1) & yes & -- &  --   \\
10613718 & 1.175880 & 12.73 & 5080 & 0.006 & 0.005 & 0.099  & 0.05(1) & 0.016(1) & -- & -- & --    \\
10991989 &  0.974475 & 10.28 & 5021 & 0.008 & 0.004 & 0.05$^7$ & 0.007(1) & 0.167(1) & -- &  -- & --    \\
11042923 &  0.390164 & 14.32 & 6086 & 0.210 & 0.208 & -- & 0.48(1) & 0.153(2) & -- & -- & --     \\
11968490 &  1.078899 & 13.70 & NA & 0.033 & 0.017 & 0.052 & 0.043(1) & 0.228(1)  & --& -- & --    \\
\enddata
\tablecomments{(1) The {\em Kepler} magnitude and effective temperature are taken from the {\em Kepler} input catalog; (2) Depths of the primary and secondary eclipses, based on our epoch-folded light curves; (3)  Eccentricity of the binary, taken from Slawson et al.~(2011) as $e_{\rm bin} = [(e\sin \omega_{\rm bin})^2+(e\cos \omega_{\rm bin})^2]^{1/2}$, except where otherwise noted; (4) Mass ratio of the two stars in the binary, $q_{\rm bin}$, and the fraction of the total {\em Kepler} luminosity contributed by the third star, $L_3/L_{\rm trip}$, as analyzed with the {\em Phoebe} binary light curve fitting code (the number in parentheses reflects the statistical uncertainty in the last significant digit(s)); (5) See Table 3 for references; (6) This object is the same as the eclipsing binary V404 Lyr (see, e.g., Pigulski et al.~2009); (7) Substituted with values from our {\em Phoebe} light curve analysis.}
\end{deluxetable}
\begin{deluxetable}{lcccccccc}
\tablewidth{0pt}
\tabletypesize{\scriptsize}
\tablecaption{\label{tab:xray} Fitted Periods, Masses, and $O-C$ Amplitudes for the Triple-star Candidates}
\tablehead{
	\colhead{Source} &
	\colhead{$P_{\rm bin}$$^1$} &
	\colhead{$P_{\rm trip} $}  & 
	\colhead{$M_3/M_{\rm trip}$} &
	\colhead{$f(M_3)$$^2$} &
	\colhead{$M_3 \sin^3 i_{\rm trip}$} &
	\colhead{$M_{\rm bin} \sin^3 i_{\rm trip}$} &
    	\colhead{$A_{\rm Roem}$$^3$} &
	\colhead{$A_{\rm phys}$$^4$} \\
	\colhead{} & 
	\colhead{(days)} &
	\colhead{(days)} & 
      	\colhead{} &
	\colhead{($M_\odot$)} &
	\colhead{($M_\odot$)} &
         \colhead{($M_\odot$)} &
	\colhead{(sec)} &
	\colhead{(sec)}
}
\startdata 
3228863 &  0.730942 & 668.4& 0.42\scriptsize{\{0.24,0.48}\} & 0.017\scriptsize{\{0.016,0.017}\} & 0.10\scriptsize{\{0.07,0.28}\} & 0.13\scriptsize{\{0.08,0.90}\} & 189\scriptsize{\{187,194\}} &    3.5\scriptsize{\{2.0,4.0\}} \\
4647652 &  1.064820 & 753.5& 0.41\scriptsize{\{0.26,0.53}\} & 0.023\scriptsize{\{0.012,0.039}\} & 0.13\scriptsize{\{0.08,0.31}\} & 0.17\scriptsize{\{0.09,0.80}\} & 228\scriptsize{\{183,274\}} &    7.5\scriptsize{\{4.7,10.4\}} \\
4909707 &  2.302370& 505.3& 0.70\scriptsize{\{0.50,0.86}\} & 0.510\scriptsize{\{0.230,1.053}\} & 1.08\scriptsize{\{0.47,2.65}\} & 0.40\scriptsize{\{0.11,2.18}\} & 493\scriptsize{\{378,627\}} &  122\scriptsize{\{81,189\}} \\
4940201 &  8.81659 & 361.6& 0.52\scriptsize{\{0.35,0.77}\} & 0.268\scriptsize{\{0.042,1.266}\} & 1.08\scriptsize{\{0.19,3.22}\} & 0.80\scriptsize{\{0.14,3.33}\} & 318\scriptsize{\{171,534\}} & 1209\scriptsize{\{846,1768\}} \\
5039441 &  2.151390 & 667.8& 0.42\scriptsize{\{0.26,0.57}\} & 0.026\scriptsize{\{0.011,0.061}\} & 0.15\scriptsize{\{0.08,0.36}\} & 0.17\scriptsize{\{0.09,0.81}\} & 220\scriptsize{\{163,293\}} &   39\scriptsize{\{24,60\}} \\
5128972 &  0.505317& 447.8& 0.55\scriptsize{\{0.38,0.69}\} & 0.094\scriptsize{\{0.079,0.108}\} & 0.29\scriptsize{\{0.20,0.66}\} & 0.23\scriptsize{\{0.09,1.08}\} & 259\scriptsize{\{244,271\}} &    3.9\scriptsize{\{2.7,4.9\}} \\
5264818 &  1.905052 & 296.3& 0.42\scriptsize{\{0.26,0.60}\} & 0.037\scriptsize{\{0.015,0.094}\} & 0.21\scriptsize{\{0.09,0.66}\} & 0.24\scriptsize{\{0.09,1.69}\} & 145\scriptsize{\{107,196\}} &   66\scriptsize{\{42,99\}} \\
5310387 &  0.441669 & 214.2& 0.16\scriptsize{\{0.10,0.20}\} & $<0.001$ & 0.03\scriptsize{\{0.02,0.07}\} & 0.15\scriptsize{\{0.09,0.55}\} &  31\scriptsize{\{ 27, 37\}} &    2.4\scriptsize{\{1.5,3.7\}} \\
5376552 &  0.503819& 334.5& 0.32\scriptsize{\{0.20,0.39}\} & 0.008\scriptsize{\{0.007,0.009}\} & 0.08\scriptsize{\{0.06,0.19}\} & 0.16\scriptsize{\{0.09,0.72}\} &  94\scriptsize{\{ 91, 98\}} &    3.3\scriptsize{\{2.0,4.1\}} \\
5384802 &  6.08309& 254.8& 0.41\scriptsize{\{0.27,0.71}\} & 0.075\scriptsize{\{0.007,0.972}\} & 0.48\scriptsize{\{0.07,2.68}\} & 0.53\scriptsize{\{0.11,2.68}\} & 165\scriptsize{\{ 75,387\}} &  754\scriptsize{\{559,1168\}} \\
5771589 & 10.74007& 113.2& 0.35\scriptsize{\{0.32,0.38}\} & 0.073\scriptsize{\{0.009,0.247}\} & 0.59\scriptsize{\{0.08,2.08}\} & 1.10\scriptsize{\{0.15,3.96}\} &  95\scriptsize{\{ 48,142\}} & 4193\scriptsize{\{3913,4493\}} \\
6370665 &  0.932316 & 285.9& 0.26\scriptsize{\{0.17,0.32}\} & 0.004\scriptsize{\{0.003,0.005}\} & 0.06\scriptsize{\{0.04,0.15}\} & 0.15\scriptsize{\{0.08,0.72}\} &  67\scriptsize{\{ 61, 74\}} &    9.0\scriptsize{\{5.7,10.9\}} \\
6525196 &  3.42060& 415.8& 0.38\scriptsize{\{0.27,0.58}\} & 0.063\scriptsize{\{0.031,0.201}\} & 0.59\scriptsize{\{0.15,1.33}\} & 0.85\scriptsize{\{0.16,3.45}\} & 215\scriptsize{\{171,318\}} &  127\scriptsize{\{91,189\}} \\
6531485 &  0.676991 &  48.3& 0.61\scriptsize{\{0.34,0.77}\} & 0.173\scriptsize{\{0.014,0.613}\} & 0.32\scriptsize{\{0.13,2.77}\} & 0.18\scriptsize{\{0.08,3.22}\} &  72\scriptsize{\{ 31,109\}} &   83\scriptsize{\{58,109\}} \\
6545018 &  3.99146 &  90.6& 0.29\scriptsize{\{0.20,0.46}\} & 0.038\scriptsize{\{0.005,0.297}\} & 0.51\scriptsize{\{0.07,1.87}\} & 1.21\scriptsize{\{0.16,3.78}\} &  66\scriptsize{\{ 33,131\}} &  572\scriptsize{\{439,866\}} \\
7289157 &  5.26640& 243.8& 0.52\scriptsize{\{0.30,0.76}\} & 0.187\scriptsize{\{0.021,1.065}\} & 0.74\scriptsize{\{0.14,2.83}\} & 0.57\scriptsize{\{0.12,2.90}\} & 218\scriptsize{\{104,387\}} &  737\scriptsize{\{504,1029\}} \\
7668648 & 27.8184 & 203.7& 0.10\scriptsize{\{0.08,0.12}\} & 0.001\scriptsize{\{$<0.001$,0.004}\} & 0.14\scriptsize{\{0.02,0.41}\} & 1.31\scriptsize{\{0.22,3.65}\} &  37\scriptsize{\{ 21, 55\}} & 4759\scriptsize{\{4097,5401\}} \\
7690843 &  0.786259 &  74.3& 0.40\scriptsize{\{0.26,0.64}\} & 0.071\scriptsize{\{0.026,0.147}\} & 0.41\scriptsize{\{0.13,1.05}\} & 0.49\scriptsize{\{0.11,2.80}\} &  71\scriptsize{\{ 51, 91\}} &   40\scriptsize{\{24,61\}} \\
7837302 & 23.83530 & 959.3& 0.44\scriptsize{\{0.26,0.73}\} & 0.177\scriptsize{\{0.017,1.281}\} & 1.03\scriptsize{\{0.15,3.37}\} & 1.13\scriptsize{\{0.16,3.74}\} & 528\scriptsize{\{244,999\}} & 2770\scriptsize{\{1748,4545\}} \\
7955301 & 15.3266& 209.5& 0.36\scriptsize{\{0.32,0.39}\} & 0.094\scriptsize{\{0.012,0.277}\} & 0.73\scriptsize{\{0.10,2.18}\} & 1.30\scriptsize{\{0.18,3.96}\} & 156\scriptsize{\{ 79,223\}} & 5788\scriptsize{\{5464,6131\}} \\
8023317 & 16.57828 & 613.5& 0.10\scriptsize{\{0.08,0.14}\} & 0.001\scriptsize{\{$<0.001$,0.007}\} & 0.10\scriptsize{\{0.02,0.42}\} & 0.85\scriptsize{\{0.17,3.33}\} &  70\scriptsize{\{ 41,131\}} &  528\scriptsize{\{ 410, 680\}} \\
8043961 &  1.559210& 476.7& 0.41\scriptsize{\{0.25,0.56}\} & 0.034\scriptsize{\{0.028,0.045}\} & 0.21\scriptsize{\{0.12,0.49}\} & 0.29\scriptsize{\{0.10,1.42}\} & 194\scriptsize{\{179,213\}} &   24\scriptsize{\{15,33\}} \\
8192840 &  0.433547& 803.9& 0.38\scriptsize{\{0.23,0.47}\} & 0.015\scriptsize{\{0.011,0.019}\} & 0.10\scriptsize{\{0.07,0.26}\} & 0.16\scriptsize{\{0.09,0.85}\} & 208\scriptsize{\{187,223\}} &    1.9\scriptsize{\{1.3,3.1\}} \\
8386865 &  1.25800 & 293.0& 0.55\scriptsize{\{0.36,0.67}\} & 0.063\scriptsize{\{0.047,0.117}\} & 0.23\scriptsize{\{0.14,0.62}\} & 0.18\scriptsize{\{0.08,1.08}\} & 171\scriptsize{\{156,210\}} &   37\scriptsize{\{26,49\}} \\
8394040 &  0.302128& 394.8& 0.71\scriptsize{\{0.47,0.84}\} & 0.353\scriptsize{\{0.287,0.414}\} & 0.70\scriptsize{\{0.50,1.58}\} & 0.28\scriptsize{\{0.10,1.81}\} & 369\scriptsize{\{345,391\}} &    5.4\scriptsize{\{3.5,7.7\}} \\
8719897 &  3.15142 & 332.7& 0.52\scriptsize{\{0.36,0.70}\} & 0.158\scriptsize{\{0.086,0.283}\} & 0.59\scriptsize{\{0.23,1.61}\} & 0.49\scriptsize{\{0.11,2.77}\} & 253\scriptsize{\{205,307\}} &  177\scriptsize{\{121,230\}} \\
8904448 &  0.865981& 548.1& 0.41\scriptsize{\{0.25,0.49}\} & 0.018\scriptsize{\{0.014,0.025}\} & 0.11\scriptsize{\{0.08,0.26}\} & 0.15\scriptsize{\{0.09,0.76}\} & 171\scriptsize{\{158,192\}} &   11\scriptsize{\{6,15\}} \\
8938628 &  6.86219& 388.1& 0.22\scriptsize{\{0.17,0.34}\} & 0.015\scriptsize{\{0.003,0.171}\} & 0.37\scriptsize{\{0.05,1.50}\} & 1.45\scriptsize{\{0.15,4.05}\} & 127\scriptsize{\{ 75,287\}} &  318\scriptsize{\{256,481\}} \\
9451096 &  1.25039& 106.7& 0.39\scriptsize{\{0.25,0.65}\} & 0.069\scriptsize{\{0.019,0.283}\} & 0.49\scriptsize{\{0.13,1.33}\} & 0.61\scriptsize{\{0.12,3.14}\} &  90\scriptsize{\{ 59,144\}} &   66\scriptsize{\{42,107\}} \\
9714358 &  6.47418 & 103.7& 0.27\scriptsize{\{0.21,0.35}\} & 0.028\scriptsize{\{0.004,0.142}\} & 0.39\scriptsize{\{0.06,1.50}\} & 1.04\scriptsize{\{0.15,3.91}\} &  65\scriptsize{\{ 35,112\}} & 1252\scriptsize{\{1041,1558\}} \\
9722737 &  0.418528& 443.9& 0.55\scriptsize{\{0.36,0.64}\} & 0.068\scriptsize{\{0.063,0.073}\} & 0.22\scriptsize{\{0.16,0.52}\} & 0.18\scriptsize{\{0.09,0.92}\} & 230\scriptsize{\{225,236\}} &    2.4\scriptsize{\{1.6,2.8\}} \\
9912977 &  0.943916 & 753.7& 0.23\scriptsize{\{0.14,0.27}\} & 0.002\scriptsize{\{0.002,0.003}\} & 0.04\scriptsize{\{0.03,0.11}\} & 0.14\scriptsize{\{0.08,0.66}\} & 105\scriptsize{\{ 94,117\}} &    3.2\scriptsize{\{1.9,4.0\}} \\
10095512 &  6.01720 & 472.6& 0.50\scriptsize{\{0.37,0.71}\} & 0.185\scriptsize{\{0.072,0.579}\} & 0.88\scriptsize{\{0.22,2.15}\} & 0.78\scriptsize{\{0.13,3.18}\} & 337\scriptsize{\{247,493\}} &  414\scriptsize{\{304,572\}} \\
10226388 &  0.660658 & 934.9& 0.60\scriptsize{\{0.39,0.72}\} & 0.124\scriptsize{\{0.101,0.150}\} & 0.35\scriptsize{\{0.23,0.83}\} & 0.24\scriptsize{\{0.09,1.30}\} & 465\scriptsize{\{434,493\}} &    3.3\scriptsize{\{2.2,4.1\}} \\
10319590 & 21.3216 & 247.1& 0.22\scriptsize{\{0.10,0.62}\} & 0.013\scriptsize{\{0.001,0.642}\} & 0.34\scriptsize{\{0.04,2.05}\} & 1.08\scriptsize{\{0.17,3.65}\} &  90\scriptsize{\{ 34,329\}} & 4193\scriptsize{\{2175,9999\}} \\
10613718 &  1.175880 &  88.1& 0.47\scriptsize{\{0.30,0.72}\} & 0.136\scriptsize{\{0.063,0.449}\} & 0.75\scriptsize{\{0.26,1.73}\} & 0.75\scriptsize{\{0.14,3.33}\} &  99\scriptsize{\{ 76,147\}} &   80\scriptsize{\{52,121\}} \\
10991989 &  0.974478 & 554.2& 0.54\scriptsize{\{0.34,0.63}\} & 0.059\scriptsize{\{0.049,0.072}\} & 0.21\scriptsize{\{0.15,0.49}\} & 0.18\scriptsize{\{0.09,0.92}\} & 256\scriptsize{\{239,274\}} &   11\scriptsize{\{7,13\}} \\
11042923 &  0.390164 & 839.0& 0.40\scriptsize{\{0.21,0.47}\} & 0.017\scriptsize{\{0.015,0.019}\} & 0.10\scriptsize{\{0.08,0.37}\} & 0.15\scriptsize{\{0.09,1.37}\} & 223\scriptsize{\{213,230\}} &    $<1$ \\
11968490 &  1.078899& 253.2& 0.63\scriptsize{\{0.43,0.80}\} & 0.333\scriptsize{\{0.287,0.387}\} & 0.88\scriptsize{\{0.55,1.69}\} & 0.52\scriptsize{\{0.14,2.20}\} & 271\scriptsize{\{256,283\}} &   38\scriptsize{\{26,49\}} \\
\enddata
\tablecomments{(1) The binary period is referenced to an epoch of BJD = 2454900; (2) Defined as $M^3_3 \sin^3 i_{\rm trip}/(M_3+M_{\rm bin})^2$, (3) See eq.~(\ref{eqn:Roemamp}) for the definition, (4) See eq.~(\ref{eqn:physamp}) for the definition. The values in curly brackets represent the 10\% lower- and 90\% upper-limits on the probability distribution. The parameter values and uncertainties reported in this table are based on $10^8$ parameter draws for a 5-day binary, and scaled proportionally to $P_{\rm bin}$.}
\end{deluxetable}
\begin{deluxetable}{lcccccccc}
\tablewidth{0pt}
\tabletypesize{\scriptsize}
\tablecaption{\label{tab:xray} Fitted Orbital Parameters for the Triple-star Candidates}
\tablehead{
	\colhead{Source} &
	\colhead{eccentricity$^1$} &
	\colhead{$\omega$$^{(2)}$}  & 
	\colhead{$\tau$$^{(3)}$} &
	\colhead{$i_m$$^4$}&
	\colhead{$v_m$$^5$} &
	\colhead{rms$^6$} &
    	\colhead{$\tau_{\rm longterm}$$^7$} &
	\colhead{Refs.} \\
	\colhead{} & 
	\colhead{} &
	\colhead{(degrees)} & 
      	\colhead{(days)} &
	\colhead{(degrees)} &
	\colhead{(degrees)} &
         \colhead{(sec)} &
	\colhead{years} &
	\colhead{}
}
\startdata 
3228863 & 0.08\scriptsize{\{0.06,0.12}\} & 209\scriptsize{\{192,224}\} &  94\scriptsize{\{ 63,123}\} & 45.4\scriptsize{\{18.4,71.5}\} &  92\scriptsize{\{ 13,139\}} &   51 & 1600 &  \\
4647652 & 0.35\scriptsize{\{0.10,0.44}\} & 184\scriptsize{\{ 42,340}\} & 459\scriptsize{\{113,644}\} & 44.9\scriptsize{\{19.5,70.4}\} &  90\scriptsize{\{ 21,160\}} &   35 &  1400&  \\
4909707 & 0.54\scriptsize{\{0.31,0.66}\} & 344\scriptsize{\{295,417}\} & 449\scriptsize{\{392,537}\} & 43.9\scriptsize{\{24.2,63.7}\} &  88\scriptsize{\{ 22,158\}} &  126 & 305 &  \\
4940201 & 0.18\scriptsize{\{0.11,0.25}\} & 163\scriptsize{\{ 42,326}\} & 319\scriptsize{\{289,340}\} & 16.3\scriptsize{\{ 9.2,21.4}\} &  54\scriptsize{\{ 18,150\}} &  167 & 41&  \\
5039441 & 0.42\scriptsize{\{0.18,0.54}\} & 187\scriptsize{\{ 36,345}\} & 336\scriptsize{\{ 48,619}\} & 45.4\scriptsize{\{24.2,66.4}\} &  87\scriptsize{\{ 21,159\}} &   39 & 566 &  \\
5128972 & 0.33\scriptsize{\{0.25,0.41}\} & 101\scriptsize{\{ 84,116}\} &  26\scriptsize{\{  7, 46}\} & 45.0\scriptsize{\{18.4,71.4}\} &  86\scriptsize{\{ 16,157\}} &   39& 1086 &  \\
5264818 & 0.37\scriptsize{\{0.13,0.53}\} & 173\scriptsize{\{ 34,332}\} & 120\scriptsize{\{ 23,270}\} & 41.4\scriptsize{\{23.2,59.0}\} &  84\scriptsize{\{ 22,154\}} &   62 & 127 &  \\
5310387 & 0.53\scriptsize{\{0.34,0.61}\} & 161\scriptsize{\{ 16,345}\} & 126\scriptsize{\{ 12,194}\} & 45.7\scriptsize{\{22.6,68.0}\} &  169\scriptsize{\{122,213\}} &   20 & 285 &  \\
5376552 & 0.40\scriptsize{\{0.35,0.45}\} & 167\scriptsize{\{161,175}\} & 302\scriptsize{\{296,309}\} & 44.3\scriptsize{\{20.5,68.4}\} &  77\scriptsize{\{ 16,171\}} &   39 & 604 &  \\
5384802 & 0.36\scriptsize{\{0.23,0.46}\} & 171\scriptsize{\{ 30,334}\} & 103\scriptsize{\{ 98,112}\} & 17.1\scriptsize{\{ 9.3,23.4}\} &  84\scriptsize{\{ 30,159\}} &  105 &  29 & 8 \\
5771589 & 0.30\scriptsize{\{0.28,0.33}\} & 214\scriptsize{\{ 38,329}\} &  75\scriptsize{\{ 74, 76}\} & 31.4\scriptsize{\{30.7,32.1}\} & 169\scriptsize{\{165,172\}} &  260 & 3.2 & 9  \\
6370665 & 0.22\scriptsize{\{0.07,0.33}\} &  92\scriptsize{\{ 15,353}\} &  291\scriptsize{\{245,396}\} & 46.3\scriptsize{\{23.1,67.7}\} &  68\scriptsize{\{ 20,140\}} &   62 & 240  &  \\
6525196 & 0.30\scriptsize{\{0.26,0.35}\} & 285\scriptsize{\{233,310}\} & 187\scriptsize{\{127,200}\} & 28.0\scriptsize{\{22.6,33.9}\} & 129\scriptsize{\{ 84,147\}} &   29 & 138  &  \\
6531485 & 0.44\scriptsize{\{0.33,0.63}\} & 315\scriptsize{\{204,347}\} &  35\scriptsize{\{ 33, 35}\} & 37.8\scriptsize{\{14.1,48.8}\} &  23\scriptsize{\{  8,175\}} &   68 & 9.5 &  \\
6545018 & 0.26\scriptsize{\{0.16,0.36}\} & 150\scriptsize{\{ 41,319}\} &  69\scriptsize{\{ 67, 71}\} & 21.8\scriptsize{\{16.8,27.7}\} &  46\scriptsize{\{ 23, 63\}} &  109 & 9 \\
7289157 & 0.36\scriptsize{\{0.27,0.47}\} & 161\scriptsize{\{ 42,320}\} &  44\scriptsize{\{ 34, 51}\} & 22.6\scriptsize{\{15.3,29.7}\} &  68\scriptsize{\{  9,172\}} &   73 & 31 & 9, 10 \\
7668648 & 0.36\scriptsize{\{0.28,0.42}\} & 185\scriptsize{\{ 40,327}\} &  29\scriptsize{\{ 20, 36}\} & 36.8\scriptsize{\{30.5,40.8}\} &  70\scriptsize{\{ 59, 81\}} & 1193 &  4 & 9, 10 \\
7690843 & 0.25\scriptsize{\{0.08,0.42}\} & 258\scriptsize{\{ 48,334}\} &  44\scriptsize{\{ 25, 59}\} & 29.1\scriptsize{\{17.1,42.2}\} & 101\scriptsize{\{ 35,149\}} &   36 & 19  &  \\
7837302 & 0.16\scriptsize{\{0.08,0.25}\} & 247\scriptsize{\{175,319}\} & 353\scriptsize{\{302,397}\} & 14.7\scriptsize{\{11.4,18.8}\} & 140\scriptsize{\{ 15,169\}} &  120 & 106 &  \\
7955301 & 0.45\scriptsize{\{0.43,0.48}\} & 161\scriptsize{\{ 36,326}\} & 187\scriptsize{\{186,188}\} & 31.6\scriptsize{\{30.8,32.4}\} & 157\scriptsize{\{153,161\}} &  326 &  8 & 9 \\
8023317& 0.23\scriptsize{\{0.18,0.29}\} & 207\scriptsize{\{ 63,336}\} & 118\scriptsize{\{ 92,145}\} & 53.0\scriptsize{\{45.8,62.4}\} &  68\scriptsize{\{ 52, 85\}} &   19 & 62 &  \\
8043961 & 0.25\scriptsize{\{0.14,0.33}\} & 192\scriptsize{\{167,212}\} & 398\scriptsize{\{363,425}\} & 34.6\scriptsize{\{16.4,54.7}\} & 102\scriptsize{\{ 11,172\}} &   50 & 400  &  \\
8192840 & 0.63\scriptsize{\{0.52,0.70}\} & 173\scriptsize{\{160,185}\} & 569\scriptsize{\{544,595}\} & 45.0\scriptsize{\{18.7,71.3}\} &  79\scriptsize{\{ 24,164\}} &   59 & 4108  &  \\
8386865 & 0.38\scriptsize{\{0.27,0.48}\} & 137\scriptsize{\{105,159}\} & 128\scriptsize{\{111,147}\} & 53.2\scriptsize{\{33.1,74.0}\} & 120\scriptsize{\{ 70,158\}} &  115 & 187 &  \\
8394040 & 0.61\scriptsize{\{0.50,0.67}\} & 123\scriptsize{\{113,131}\} & 296\scriptsize{\{288,305}\} & 43.8\scriptsize{\{17.8,70.8}\} &  73\scriptsize{\{ 19,159\}} &   96 & 1088  &  \\
8719897 & 0.24\scriptsize{\{0.13,0.31}\} & 291\scriptsize{\{267,317}\} &  90\scriptsize{\{ 68,103}\} & 17.4\scriptsize{\{ 9.2,25.2}\} &  98\scriptsize{\{ 29,151\}} &   51 & 96  &  \\
8904448 & 0.59\scriptsize{\{0.50,0.66}\} & 135\scriptsize{\{125,143}\} & 443\scriptsize{\{431,454}\} & 40.1\scriptsize{\{18.3,63.9}\} &  68\scriptsize{\{ 12,166\}} &   32 & 950 &  \\
8938628 & 0.31\scriptsize{\{0.26,0.35}\} & 282\scriptsize{\{221,327}\} & 339\scriptsize{\{314,348}\} & 17.4\scriptsize{\{12.4,21.1}\} & 133\scriptsize{\{ 27,160\}} &   21 & 60  &  \\
9451096 & 0.24\scriptsize{\{0.10,0.36}\} & 183\scriptsize{\{ 53,313}\} &  60\scriptsize{\{  8, 97}\} & 23.4\scriptsize{\{11.9,37.1}\} &  91\scriptsize{\{ 33,150\}} &   19 & 25  &  \\
9714358 & 0.26\scriptsize{\{0.20,0.32}\} & 154\scriptsize{\{ 29,329}\} &  77\scriptsize{\{ 76, 78}\} & 16.8\scriptsize{\{13.8,20.8}\} & 134\scriptsize{\{120,149\}} &  131& 4.6  & 9 \\
9722737 & 0.22\scriptsize{\{0.16,0.27}\} &  29\scriptsize{\{ 14, 46}\} & 424\scriptsize{\{416,461}\} & 45.1\scriptsize{\{18.4,71.8}\} &  229\scriptsize{\{160,242\}} &   48 & 1290  &  \\
9912977 & 0.31\scriptsize{\{0.16,0.39}\} & 251\scriptsize{\{213,301}\} & 260\scriptsize{\{187,359}\} & 45.0\scriptsize{\{18.7,71.2}\} & 103\scriptsize{\{ 24,159\}} &   22 & 1650  &  \\
10095512 & 0.18\scriptsize{\{0.12,0.23}\} &  67\scriptsize{\{ 37,101}\} & 442\scriptsize{\{420,480}\} & 13.6\scriptsize{\{ 6.9,18.7}\} &  89\scriptsize{\{ 28,150\}} &   23 & 100 &  \\
10226388 & 0.32\scriptsize{\{0.24,0.39}\} & 281\scriptsize{\{263,300}\} & 755\scriptsize{\{713,797}\} & 44.9\scriptsize{\{18.4,71.9}\} &  80\scriptsize{\{ 21,158\}} &  101 & 3588  &  \\
10319590 & 0.14\scriptsize{\{0.05,0.32}\} & 182\scriptsize{\{ 39,327}\} &  95\scriptsize{\{ 82,111}\} & 10.4\scriptsize{\{ 6.6,21.3}\} & 102\scriptsize{\{ 11,171\}} &  470 &  7.8 & 9 \\
10613718 & 0.18\scriptsize{\{0.05,0.29}\} & 240\scriptsize{\{138,291}\} &  20\scriptsize{\{  7, 76}\} & 18.1\scriptsize{\{ 9.7,29.0}\} & 121\scriptsize{\{ 26,157\}} &   66 & 18 &  \\
10991989 & 0.30\scriptsize{\{0.21,0.37}\} & 189\scriptsize{\{178,202}\} &  571\scriptsize{\{553,592}\} & 43.0\scriptsize{\{18.9,68.5}\} & 128\scriptsize{\{ 21,165\}} &   82 & 861  &  \\
11042923 & 0.17\scriptsize{\{0.09,0.25}\} &  34\scriptsize{\{-16,55}\} & 679\scriptsize{\{587,747}\} & 45.3\scriptsize{\{19.6,71.2}\} &  92\scriptsize{\{ 25,162\}} &   57 & 4950  &  \\
11968490 & 0.40\scriptsize{\{0.31,0.46}\} & 117\scriptsize{\{107,127}\} & 216\scriptsize{\{209,224}\} & 32.6\scriptsize{\{16.2,48.9}\} &  57\scriptsize{\{ 29,128\}} &   43 & 162  &  \\
\enddata
\tablecomments{(1) Orbital eccentricity of the triple system; (2) longitude of periastron of the orbit of the triple system (specifically $\omega$ describing the binary CM); (3) time of periastron passage of the triple system; (4) mutual inclination angle between the orbital planes of the binary and triple; (5) angle between the triple's periapse and the plane of the binary (see Fig.~\ref{fig:geometry}) -- $v_m$ runs between $0^\circ$ and $180^\circ$ because of the way it appears in eq.~(\ref{eqn:phys2}); (6) rms scatter of the $O-C$ points about the best-fitting model; (7) timescale for longer-term perturbations in the triple system calculated here simply as $P_{\rm trip}^2/P_{\rm bin}$ (see eq.~\ref{eqn:longterm}); (8) Fabrycky (2010); (9) Slawson et al.~(2011); (10) Carter et al.~(2013).  The values in curly brackets represent the 10\% lower- and 90\% upper-limits on the probability distribution.  The parameter values and uncertainties reported in this table are based on $10^8$ parameter draws for a 5-day binary, and scaled proportionally to $P_{\rm bin}$.}
\end{deluxetable}

\end{document}